\documentclass{aa}

\usepackage{graphicx}
\usepackage{xcolor}
\usepackage{txfonts}
\usepackage[utf8]{inputenc}
\usepackage{multirow}

\begin{document}

\title{Mass density profiles at kiloparsec scales using \\ the sub-millimetre galaxies magnification bias}
\titlerunning{kpc-scale mass density profiles with Magnification Bias}
\authorrunning{Crespo D. et al.}

\author{Crespo D.\inst{1,2}, Gonz{\'a}lez-Nuevo J.\inst{1,2}, Bonavera L.\inst{1,2}, Cueli M. M.\inst{3,4}, Casas J. M.\inst{1,2}}

\institute{$^1$Departamento de Fisica, Universidad de Oviedo, C. Federico Garcia Lorca 18, 33007 Oviedo, Spain\\
$^2$Instituto Universitario de Ciencias y Tecnologías Espaciales de Asturias (ICTEA), C. Independencia 13, 33004 Oviedo, Spain\\
 $^3$SISSA, Via Bonomea 265, 34136 Trieste, Italy\\
$^4$IFPU - Institute for fundamental physics of the Universe, Via Beirut 2, 34014 Trieste, Italy\\         
}
\date{Received xxx, xxxx; accepted xxx, xxxx}
\abstract
{Gravitational lensing is a powerful tool for studying the distribution of mass in the Universe. Understanding the magnification bias effect in gravitational lensing and its impact on the flux of sub-millimetre galaxies (SMGs) is crucial for accurate interpretations of observational data.
}
{This study aims to investigate the magnification bias effect in the context of gravitational lensing and analyse the mass density profiles of different types of foreground lenses, including quasi-stellar objects (QSOs), galaxies, and galaxy clusters. The specific goals are to compare the lens types, assess the impact of angular resolution on the analysis, and determine the adequacy of theoretical mass density profiles in explaining the observed data.
}
{The magnification bias was estimated using the cross-correlation function between the positions of background SMGs and foreground lens samples. Stacking techniques were employed to enhance the signal at smaller angular separations, and the more precise positions from the WISE catalogue were utilised to improve positional accuracy. Four different theoretical mass density profiles were analysed to extract additional information.
}
{The cross-correlation measurements revealed distinctive central excess and outer power-law profiles, with a lack of signal in the intermediate region. The lens types exhibited varying signal strengths, with QSOs producing the strongest signal and galaxy clusters showing weaker signals. The analysis of mass density profiles indicated limitations in the selected profiles' ability to explain the observed data, highlighting the need for additional considerations. The lack of extended emission in the QSO sample suggested possible influences from close satellites along the line of sight in the other lens types.
}
{The study provides valuable insights into the magnification bias effect and mass density profiles in gravitational lensing. The results suggest the presence of isolated galactic halos and the importance of considering environmental factors and close satellites in future investigations. The derived masses and best-fit parameters contribute to our understanding of lensing systems and provide constraints on the nature of central galaxies. Notably, the intriguing lack of signal around 10 arcsec challenges current understanding and calls for further quantitative analysis and confirmation of the observed feature.
}


\maketitle


\section{Introduction}

Magnification bias is a gravitational lensing effect that consists of an increase, or decrease, in the detection probability of background sources near the position of lenses, producing a modification of the background source number counts, $n(>S) = n_0\,S^{-\beta}$. This implies an excess of background objects produced by magnification bias when $\beta>1$ \citep{BON22}. In particular, in this work, we are interested in the case of very steep source number counts, $\beta>2$, which is when the magnification bias effect produces a significant boost in the number of detected background sources.
For that reason, the objects used as background sources in this study are sub-millimetre galaxies (SMGs), which are considered an optimal background sample for magnification bias studies thanks to their steep source number counts, $\beta \sim 3$, and their high redshift, $z>1$ \citep{GON14,GON17}.

The magnification bias effect can be quantified using the angular cross-correlation function (CCF). This function compares the expected excess between two sets of source samples with different redshift distributions, while taking into account the absence of magnification. Several studies \citep{Scr05,Men10,Hil13,Bar01} have employed this method. In particular, 
magnification bias analyses with SMGs explored the projected mass density profile and concentration of foreground samples of quasi-stellar objects \citep[QSOs][]{BON19}, have been utilised in cosmological studies \citep{BON20, GON21, BON21} and have also provided observational constraints on the halo mass function \citep{CUE21, CUE22}.
Moreover, \citet{DUN20} directly measured this effect with the Atacama Large Millimetre Array (ALMA).

QSOs, galaxies, and galaxy clusters can all serve as potential foreground lenses, as in this study.
QSOs are highly luminous active galactic nuclei (AGNs) that are suitable for lensing studies due to their detectability across a wide range of distances. They are usually employed as background sources \citep[e.g.][]{Ogu06,Ogu08, COU10, COU12, Har15, Dan17} and, in particular to study magnification bias \citep[e.g.,][]{Bar93,Scr05,Men10}. However, authors such as \cite{Man09} and \cite{Luo22} have measured weak-lensing shear distortions using AGNs as lenses.

Large galaxy surveys have proven to be highly valuable in cosmological studies \citep[e.g. the Baryon Oscillation Spectroscopic Survey,][]{YOR00Y, DAW13}. The potential of measuring lensing-induced cross-correlations between SMGs and low-redshift galaxies was first explored by \cite{WAN11}, who provided convincing evidence of this effect. Subsequently, \cite{GON14, GON17} conducted more comprehensive studies using the CCF between SMGs and massive galaxies. More recently, the magnification bias on SMGs caused by massive optical galaxies at $z<<1$ has been analysed to derive complementary and independent constraints on the main cosmological parameters \citep{BON20, GON21, BON21}. In addition, the magnification bias of the SMGs using both QSOs and galaxies as lenses has been used in \cite{CRE22} to study and compare their mass density profiles and estimate their masses and concentrations.

Galaxy clusters serve as massive bound systems that allow us to study the large-scale structure of the Universe, often found at the intersections of filamentary structures. They are valuable for cosmological investigations, such as exploring galaxy evolution \citep[e.g.,][]{DRE80,BUT78,BUT84,GOT03} and studying lensed high-redshift galaxies \citep[e.g.,][]{BLA99}. The correlation between galaxy clusters and selected background objects has been utilised to investigate potential lensing effects \citep[e.g.,][]{MYE05,LOP08}, including CCF measurements using SMGs as background samples to estimate the masses and concentration of the galaxy clusters \citep[][]{FER22}.

While weak lensing events produce fainter signals compared to strong lensing events, they occur more frequently, making them suitable for study using stacking techniques. Stacking involves combining the signals of weak or undetectable objects to enhance the overall signal and reduce background emission, enabling the extraction of statistical properties. Stacking has been applied in various studies, including recovering the integrated signal of the Sachs-Wolfe effect using \textit{Planck} data \citep{Pla14,Pla16b} and examining the faint polarised signal of radio and infrared sources \citep[see][]{Stil14, BON17a, BON17b}.
Additionally, stacking techniques have been used to determine the mean spectral energy distribution (SED) of optically selected quasars \citep{Bia19} and recover the weak gravitational lensing signal in the cosmic microwave background through the \textit{Planck} lensing convergence map \citep{Bia18}. Stacking has also been employed to investigate star formation in dense environments of lensing haloes at $z\sim 1$ \citep{Wel16}, and more recently  to recover the CCF signal between SMGs and galaxy clusters of different richness ranges \citep{FER22} and between SMGs, QSOs and galaxies \citep{CRE22}.

This study makes use of the stacking technique to estimate and compare the mass density profiles at kiloparsec scales of three distinct types of lenses: QSOs, galaxies, and galaxy clusters. The organisation of the paper is as follows: Section \ref{sec:data} provides details on the data used, Section \ref{sec:method} describes the methodology applied and presents the obtained measurements, Section \ref{sec:discuss} discusses the adopted theoretical mass density profiles and analyses the data using these profiles, and finally, Section \ref{sec:concl} summarises the conclusions of this study. Throughout the paper, the cosmological model adopted is a flat $\Lambda$ cold dark matter ($\Lambda$CDM) model with cosmological parameters estimated by \cite{PLA18_VI} as $\Omega_m = 0.31$, $\sigma_8 = 0.81$, and $h = H_0 / 100$ km s$^{-1}$ Mpc$^{-1} = 0.67$.


\begin{figure}[htp]
  \centering
    \includegraphics[width=\linewidth]{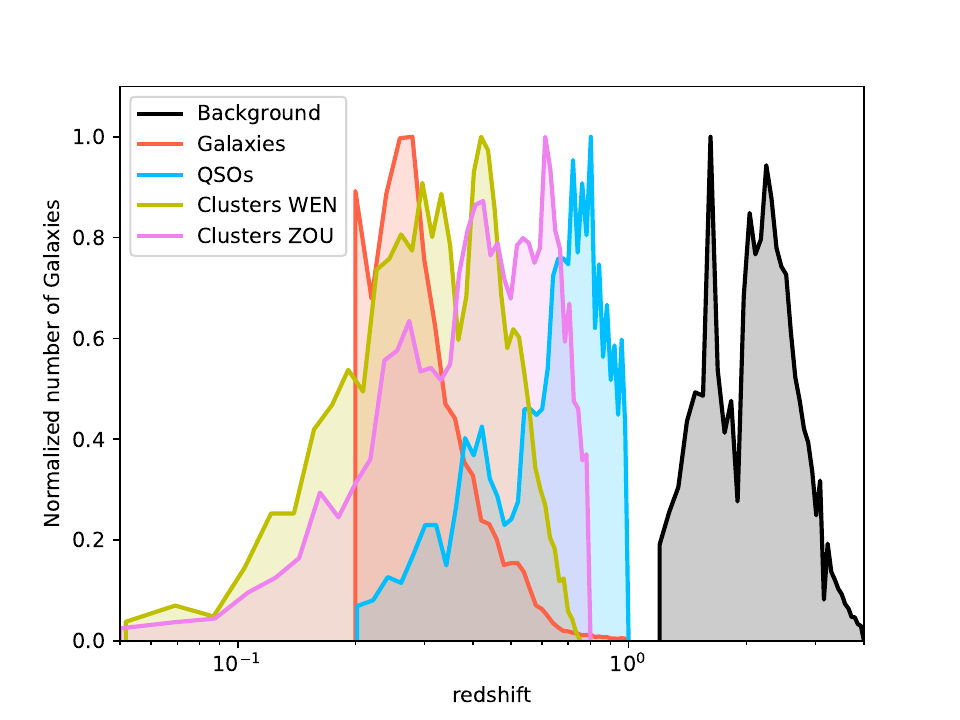}
  \caption{Redshift distribution of the four different types of foreground lenses and background sources selected in WISE (in black). The lens samples are the galaxies (red), the QSOs (cyan), the WEN clusters (gold) and the ZOU clusters (purple).}
  \label{fig:zdistro_hist}
\end{figure}

\section{Data}
\label{sec:data}
\subsection{Foreground samples}

In this work we conducted a comprehensive analysis of three distinct kinds of lenses by employing four separate catalogues. These catalogues consist of a galaxy sample, a QSO sample and two cluster samples, which provided us with a diverse range of lens masses for the analysis. 
The samples were selected in order to overlap spatially (at least partially) with the officially detected galaxies from the Herschel Astrophysical Terahertz Large Area Survey (H-ATLAS) data. These observations were obtained through the Herschel Space Observatory and cover a sky area of approximately 610 square degrees across five distinct fields. Three of these fields, known as the Galaxy and Mass Assembly (GAMA) fields or Data Delivery 1 (DR1), are located on the celestial equator at 9, 12, and 14.5 hours (G9, G12 and G15). The remaining fields, referred to as Data Delivery 2 (DR2) or the North and South Galactic Poles (NGP and SGP), cover a total area of 180.1 and 317.6 square degrees, respectively.

The galaxy sample was obtained from the GAMA II survey \citep{DRI11,BAL10,BAL14} and had already been used in previous CCF studies at larger angular scales, conducted for example by \citet{GON17, GON21} and \cite{CRE22}, the latter with the stacking technique. 
The GAMA II survey was carried out in conjunction with the H-ATLAS survey. This collaboration aimed to maximise the overlapping area between the two surveys, covering three equatorial regions at 9, 12, and 14.5 hours. 
In this study, we used the GAMA fields and the combined area of overlap amounted to approximately 207 square degrees (only 144 square degrees were used in our analysis). The foreground galaxy sample specifically included GAMA II sources with redshifts ranging from 0.2 to 0.8 as in previous works, totalling around 102672 galaxies. Their average spectroscopic redshift was measured at $\left\langle z\right\rangle =0.3^{+0.1}_{-0.1}$. The distribution of their redshifts is depicted by the red histogram in Fig. \ref{fig:zdistro_hist}. 

The QSO sample was derived from the dataset used in the study by \cite{BON19} and afterwards by \cite{CRE22}. This particular QSO sample was obtained from the Sloan Digital Sky Survey (SDSS), specifically the SDSS-II and SDSS-III Baryon Oscillation Spectroscopic Survey (BOSS) catalogues, encompassing an area of 9376 square degrees. We employed the 7th \citep[DR7]{SCH10} and 12th \citep[DR12]{PAR17} data releases of SDSS \citep[see][for a detailed description of the QSO target-selection process]{ROS12}. The selection process primarily relied on the DR7 catalogue, which mainly included QSOs at lower redshifts ($z<2.5$). On the other hand, the DR12 sample specifically targeted QSOs at higher redshifts ($z>2.5$), resulting in a secondary peak around $z\sim0.8$ due to colour degeneracy in the photometric data selection process. Approximately 4\% of the DR7 objects with $z > 2.15$ were re-observed for DR12, leading to the creation of a combined sample where any QSO present in both DR7 and DR12 was included only once.

To minimise potential overlap and contamination between the foreground and background samples in terms of redshift, we restricted our selection to QSOs with redshifts ranging from $z=0.2$ to $z=1.0$. This criterion resulted in a total of 1649 QSOs in the shared area with the background sample (the GAMA fields and the NGP). The redshift distribution of the selected QSOs is illustrated in Fig. \ref{fig:zdistro_hist} using a cyan histogram, with a mean redshift of $\left\langle z\right\rangle =0.7^{+0.1}_{-0.2}$ (the uncertainty indicates the 1$\sigma$ limits).

Regarding the galaxy clusters, we obtained one of the samples from the catalogue presented in \cite{WEN12}. This catalogue comprises a total of 132684 galaxy clusters from SDSS-III and provides photometric redshift information for clusters in the range of $0.05\leq z< 0.8$. Specifically, we selected the objects corresponding to the NGP region and the three H-ATLAS GAMA fields. This selection process resulted in a sample of 3598 galaxy clusters, which serve as our target lenses. In Fig. \ref{fig:zdistro_hist}, the gold histogram represents the redshift distribution of the foreground sources, with a mean redshift of $\langle z\rangle=0.38^{+0.23}_{-0.22}$. This sample is also used in \cite{FER22}, where they perform a richness analysis. 

We expanded our analysis with clusters as lenses by incorporating a catalogue of galaxy clusters extracted also from the SDSS, as is presented in the study by \cite{ZOU21}. This updated catalogue includes a significantly larger number of clusters, totalling 540432 clusters, located at redshifts lower than 1 within the DESI legacy imaging surveys \citep{DESI}. 
These surveys span an extensive sky area of approximately 20000 square degrees. The clusters' average mass measures is around $1.23 \times 10^{14} M_\odot$. Considering only those targets in the GAMA fields and NGP yielded 9056 clusters to be used for the analysis.
This expanded sample size, approximately three times larger than the previous cluster catalogue, is expected to yield more robust and statistically significant results. Incorporating both cluster catalogues in our analysis is an opportunity for a comparative analysis and to determine if it leads to more accurate and reliable outcomes.
The purple histogram in Fig. \ref{fig:zdistro_hist} shows the redshift distribution of this sample of foreground sources, with a mean redshift of $\langle z\rangle=0.498^{+0.24}_{-0.30}$. The errors were determined by considering the 95\% confidence intervals derived from the redshift distribution of the sample.

Table \ref{tab:lenses_sum} summarises the information of the four foreground lens samples. From left to right, the columns are: the lens sample, the number of sources included in the sample, the mean redshift of the sample, and the number of sources in each considered field.

\begin{table}[htp]
\begin{tabular}{c c c c c c c}
\hline
\hline
Sample & Sources & $\langle z \rangle$ & G09   & G12   & G15 & NGP\\ \hline

\small{Galaxies} & 102672 &0.304 & 35189 & 33768 & 33715 & 0    \\ \hline
\small{QSOs} & 1649 & 0.723 & 322   & 295   & 315   & 717  \\ \hline
\small{\begin{tabular}[c]{@{}c@{}}Clusters\\ WEN\end{tabular}} & 3598 & 0.382& 574   & 588   & 613   & 1823 \\ \hline
\small{\begin{tabular}[c]{@{}c@{}}Clusters\\ ZOU\end{tabular}} & 9056 & 0.498 & 1490  & 1412  & 1550  & 4604 \\ \hline
\end{tabular}
\caption{Summary of the foreground lens samples. From left to right: the name of the lens sample, the total number of sources included in the sample, the mean redshift of the sample redshift distribution and the number of sources in each considered field.}
\label{tab:lenses_sum}
\end{table}

\subsection{Background sample}

The background sample selection was performed similarly to previous works \citep{GON17,GON21, BON19, BON20,FER22, CRE22} on the officially detected galaxies from Herschel Astrophysical Terahertz Large Area Survey (H-ATLAS) data collected by the \textit{Herschel Space Observatory} \citep{PIL10}. 
The H-ATLAS survey employs two instruments, namely the Photodetector Array Camera and Spectrometer (PACS) and the Spectral and Photometric Imaging REceiver (SPIRE). These instruments operate in five photometric bands: 100, 160, 250, 350, and 500 micrometers ($\mu$m). In both H-ATLAS data releases, a 4$\sigma$ detection limit at 250 $\mu$m was implicitly applied, corresponding to a flux density of approximately 29 mJy. The 1$\sigma$ noise for source detection, which includes both confusion and instrumental noise, was measured at 7.4 mJy at 250 $\mu$m. Additionally, a 3$\sigma$ limit at 350 $\mu$m was employed, following a similar approach as in previous studies, to enhance the reliability of photometric redshift estimation.

In this research, for the purpose of improving the positional accuracy of the background sample, we made use of the NASA's Wide-field Infrared Survey Explorer \citep[WISE;][]{WIS10} All-Sky Data Release. The WISE mission mapped the entire sky in four infrared wavelengths (3.4, 4.6, 12, and 22 $\mu m$, referred to as W1, W2, W3 and W4) and created an atlas of the universe. SMGs emit mainly between approximately 200 and 1000 microns but WISE observed the sky in the four previous infrared bands, so that means that WISE may not be able to detect some of the SMGs directly, but it can still play a role in studying these objects when used in combination with H-ATLAS observing at different wavelengths. 
In order to improve the positional accuracy of the SMGs, we performed a cross match with the WISE catalogue so that we could use the WISE position for the CCF measurement. We adopted as the search radius the FWHM of H-ATLAS at $250 \mu m$, that is, $17.6$ arcsec, and we discarded the counterparts with W2-W3<2, typically corresponding to elliptical galaxies.

Since in WISE we often have more than one counterpart associated with an SMG position in H-ATLAS, we performed the cross-match in three different ways: by selecting the nearest and farthest sources from the H-ATLAS position as well as one at a random distance from it. As we did not discern significant differences between the three cross-matches, we used the nearest source position option. The WISE positional accuracy is well described by a Gaussian distribution with a $\sigma=0.3$ arcsec \citep{WIS10}, which is eight times more accurate than that of the H-ATLAS positions ($\sigma=2.4$ arcsec).
The black histogram in Fig. \ref{fig:zdistro_hist} depicts the redshift distribution of the background sources, with a mean redshift of $\langle z\rangle=2.25^{+0.95}_{-0.91}$.

\begin{figure}[htp]
  \centering
\includegraphics[width=0.5\textwidth]{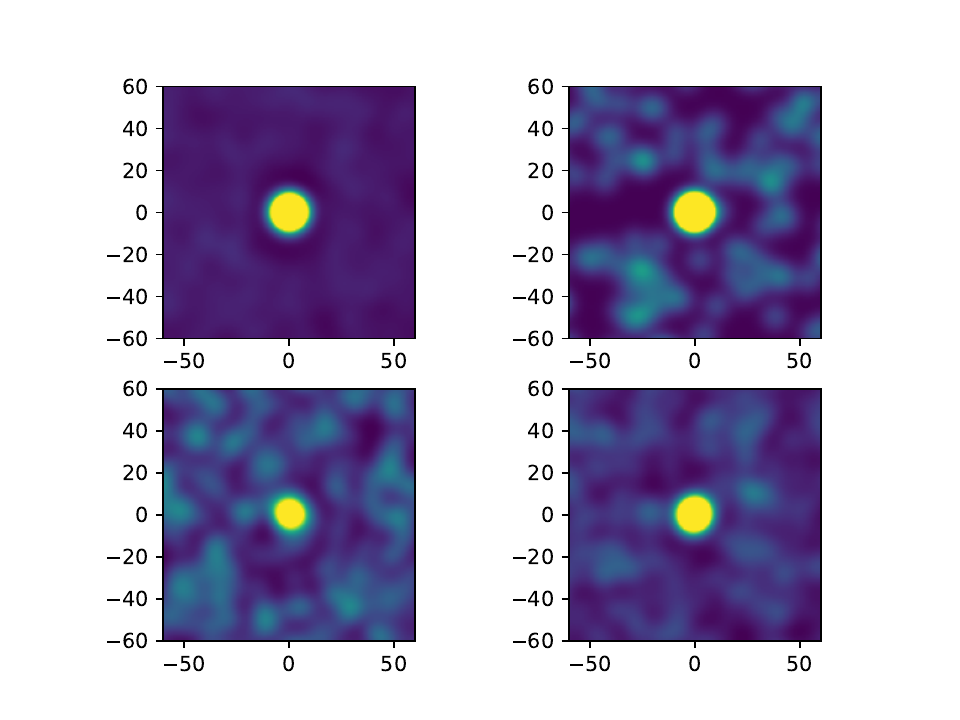}
  \caption{Staking maps obtained for the four different foreground samples. Top panel: using galaxies on the left and QSOs on the right. Bottom panel: using WEN clusters on the left and ZOU on the right. The pixel size is 0.5 arcsec and the applied smoothing is $\sigma$=2.4 arcsec.}
  \label{fig:stackingmaps}
\end{figure}

\section{Measurements}
\label{sec:method}
\subsection{Stacking}
\label{sec:stack}

The stacking technique is a valuable method used to estimate the average flux density of a large collection of sources that are individually too faint to be effectively analysed \citep[][]{Dol06,Mar09,Bet12}. This approach is particularly useful in our scientific context as it allows us to obtain reliable statistical information even when the signal is obscured by noise, making direct detection unfeasible. Typically, the stacking technique involves aggregating specific patches of the sky to enhance the underlying signal, which would otherwise remain undetectable in isolated instances. This technique enables not only statistical measurements of the flux density but also other related analyses.

Previously, a slightly modified version of this method was employed by \citet{BON19}, \citet{FER22} and \citet{CRE22} to investigate the CCF signal arising from the magnification bias between different types of lenses. These studies performed stacking analyses not on the flux densities of the sources but rather on their positions. This is because the primary focus of interest lies in the number of background sources in proximity to the positions of the lenses. For example, \cite{BON19} examined the stacked magnification bias of lensed SMGs at positions indicated by QSOs, while \cite{FER22} investigated the stacked magnification bias resulting from galaxy clusters acting as lenses on background SMGs. In \cite{CRE22}, the objective was to leverage the magnification bias exhibited by SMGs using both QSOs and galaxy lens samples to examine the mass density profiles of these foreground samples and estimate their respective masses and concentrations. By stacking the positions of the sources, these studies extracted valuable insights into the magnification effects induced by the foreground lenses.

The basic idea was to search for background sources in a circular area centred on the lens position. The positions of the pairs were recorded on a square map with a specific pixel size and number of pixels, determined by the search radius and desired angular resolution (limited by the positional accuracy of the catalogues). The maps obtained for all the lenses were then aggregated and normalised to the total number of lenses, producing the final stacked map. To account for the positional accuracy of the catalogues, a Gaussian Convolution kernel with a standard deviation corresponding to the positional accuracy of the background sample was applied: $\sigma=2.4$ arcsec for the H-ATLAS catalogue \citep[SMGs;][]{BOU16,MAD18} and $\sigma=0.3$ arcsec for the WISE positions \citep{WIS10}.

A zoomed-in view of the central part of the resulting maps for all the lens samples is shown in Fig. \ref{fig:stackingmaps}. The panels have the same colour scale and resolution, allowing for preliminary comparisons between the lens samples. Notably, there is an empty region around the central excess in all cases, and the diffuse background is much fainter in the case of galaxies compared to the other samples. Additionally, the size of the central part in the WEN cluster samples appears smaller than in the other samples.

To estimate the CCF from the maps, we applied the same procedure described in \cite{CRE22}, with a slight difference in the random part of the cross-correlation estimation provided by \cite{Dav83}:
\begin{equation}
    \tilde{w}_{x}(\theta)=\frac{\text{DD}(\theta)}{\text{RR}(\theta)}-1
    \label{eq:wx}
,\end{equation}
where the measurements were obtained by creating a series of annuli with their centres positioned at the centre of the map. The annuli have increasing radii that grow logarithmically by 0.05, starting from 1 arcsecond. The pixel values within each annuli are summed up to calculate DD and then normalised to the total. Instead of repeating the same process for the random map to generate Random-Random (RR) as done in \cite{CRE22}, we computed the RR term theoretically from the annulus area considering a constant surface density.

To compute the CCF uncertainty, the first four rings were divided into sections of equal area. However, the subsequent rings had sufficient pixels to allow for division into 15 sections of equal area. Following the approach described in \cite{BON19}, a jackknife method was applied within each annulus to estimate the uncertainties for DD. These uncertainties were then propagated using \eqref{eq:wx} to determine the uncertainty. 

\begin{figure}[ht]
  \centering
\includegraphics[width=\linewidth]{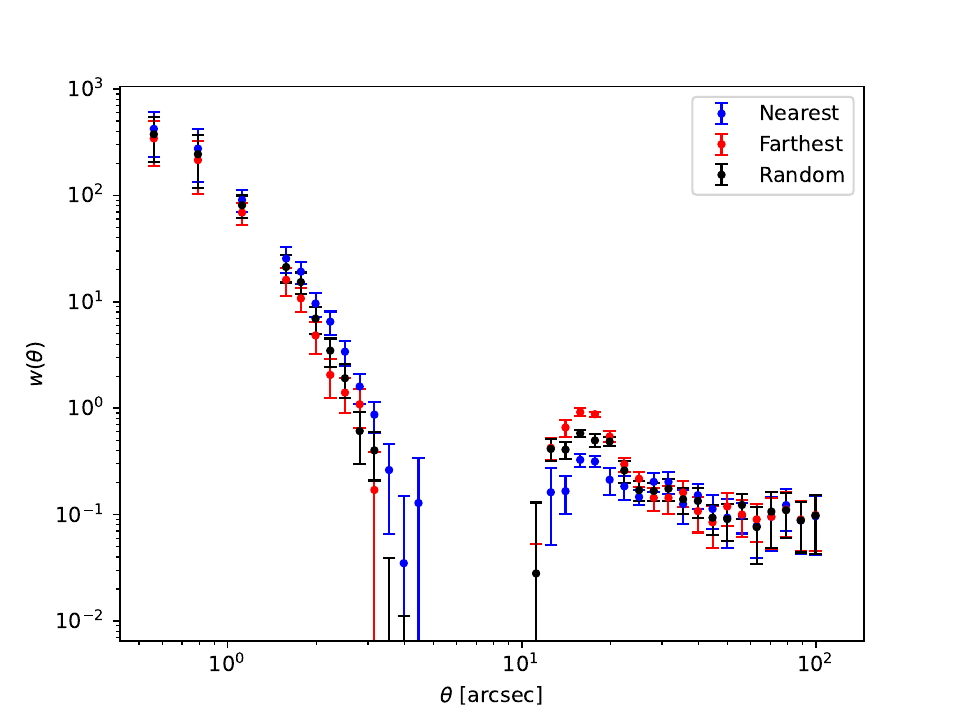}
  \caption{Cross-correlation data obtained with staking and a smoothing of $\sigma$=2.4 arcsec. The data were obtained using the galaxies foreground lens samples and the SMGs as the background sample using the WISE positions. The three cross-match approaches are depicted in blue (adopting the nearest counterpart), red (the farthest counterpart) and black (picking randomly among the possible counterparts).}
  \label{fig:xmatch}
\end{figure}

Figure \ref{fig:xmatch} represents the CCF measured for the galaxy sample obtained for three different cross-match approaches between HATLAS and WISE sources: by choosing the nearest and farthest counterparts in WISE, and a farthest and random counterpart in WISE, in blue, red, and black respectively. As can be noted, the measurements are very similar, all three cases showing a lack of signal between 5-10 arcsec. It seems that the exact cross-matching approach does not affect our main results, and therefore we decided to adopt the nearest counterpart approach for the analysis in this work. 

\subsection{Results}

\begin{figure*}[htp]
 \centering
 \includegraphics[width=0.4\textwidth]{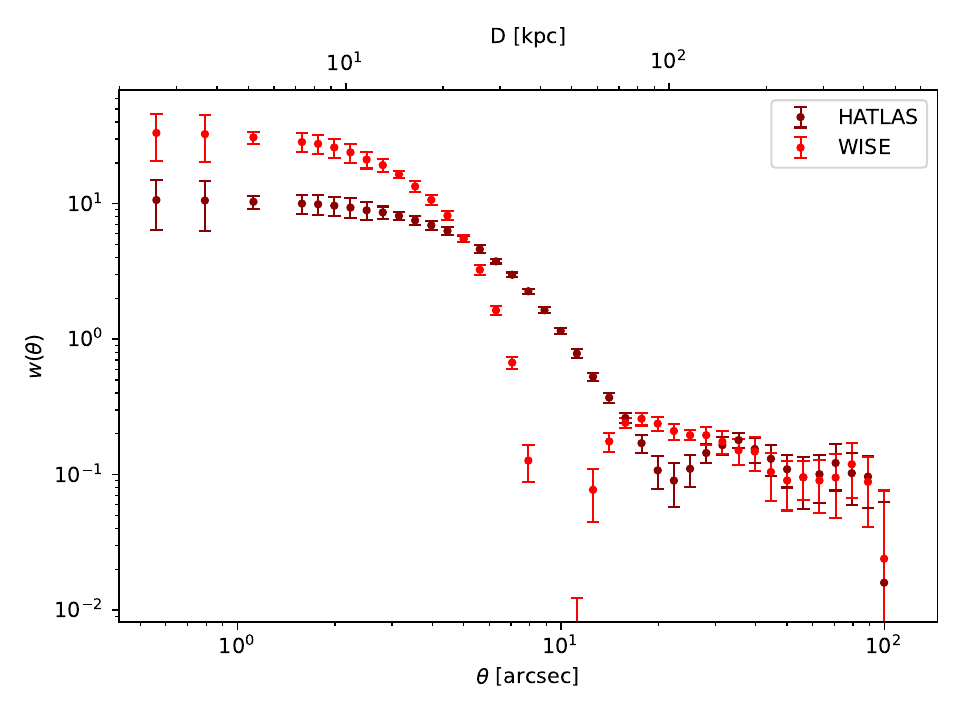}
 \includegraphics[width=0.4\textwidth]{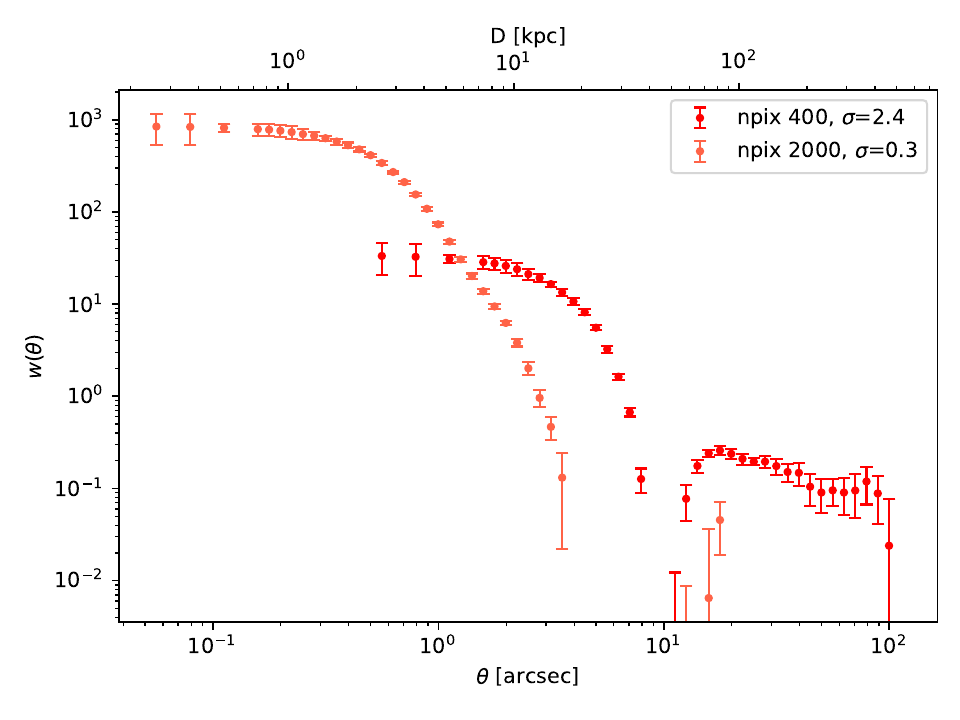}\\
  \includegraphics[width=0.4\textwidth]{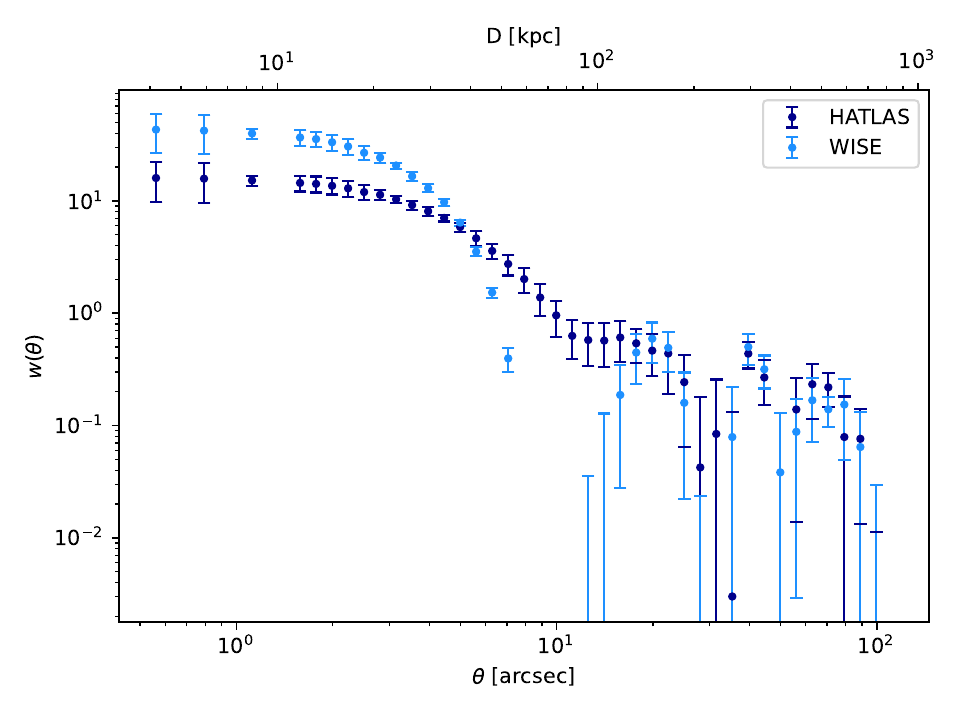}
  \includegraphics[width=0.4\textwidth]{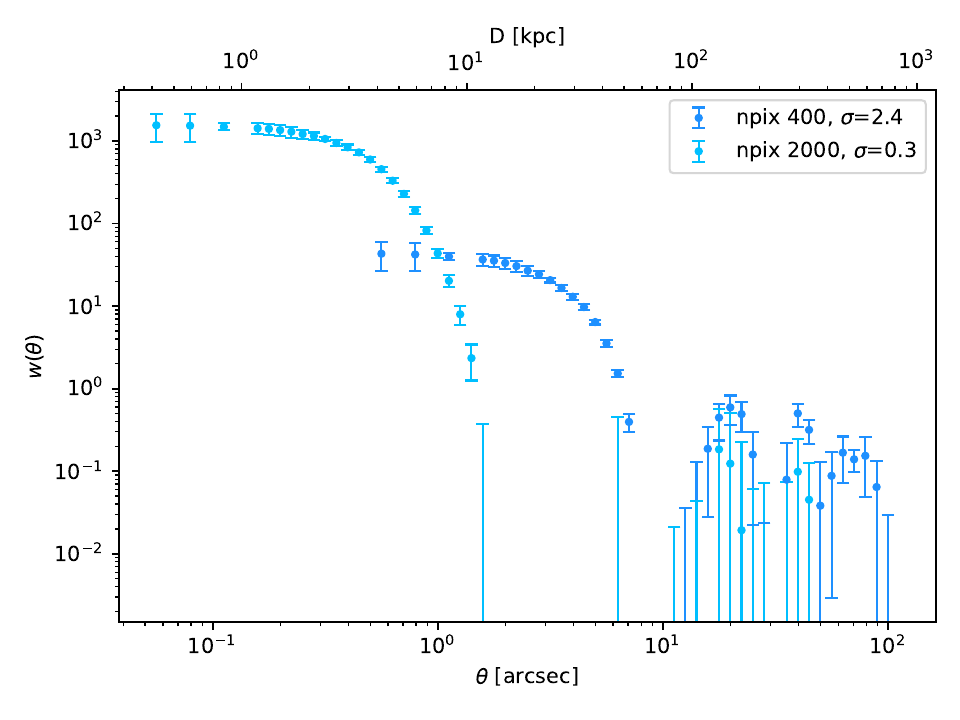}\\
 \includegraphics[width=0.4\textwidth]{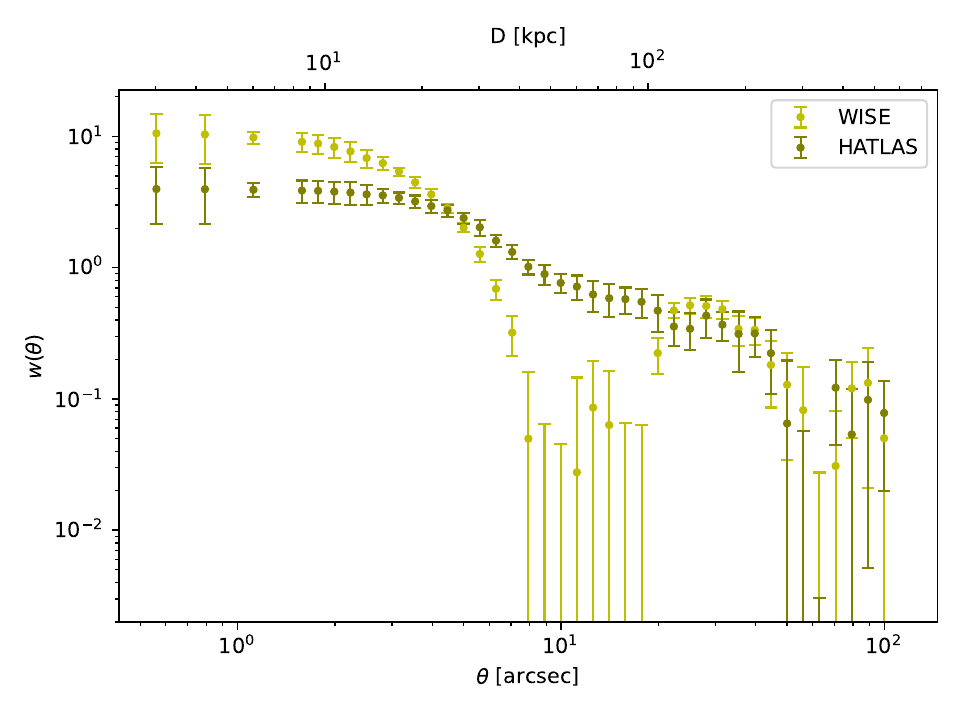}
 \includegraphics[width=0.4\textwidth]{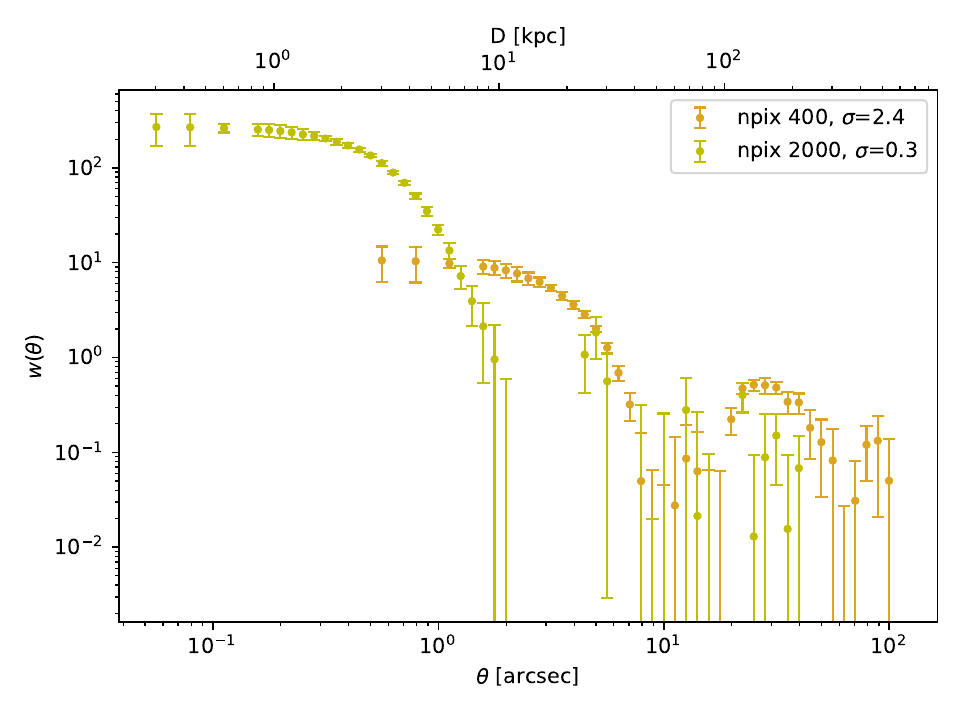}\\
 \includegraphics[width=0.4\textwidth]{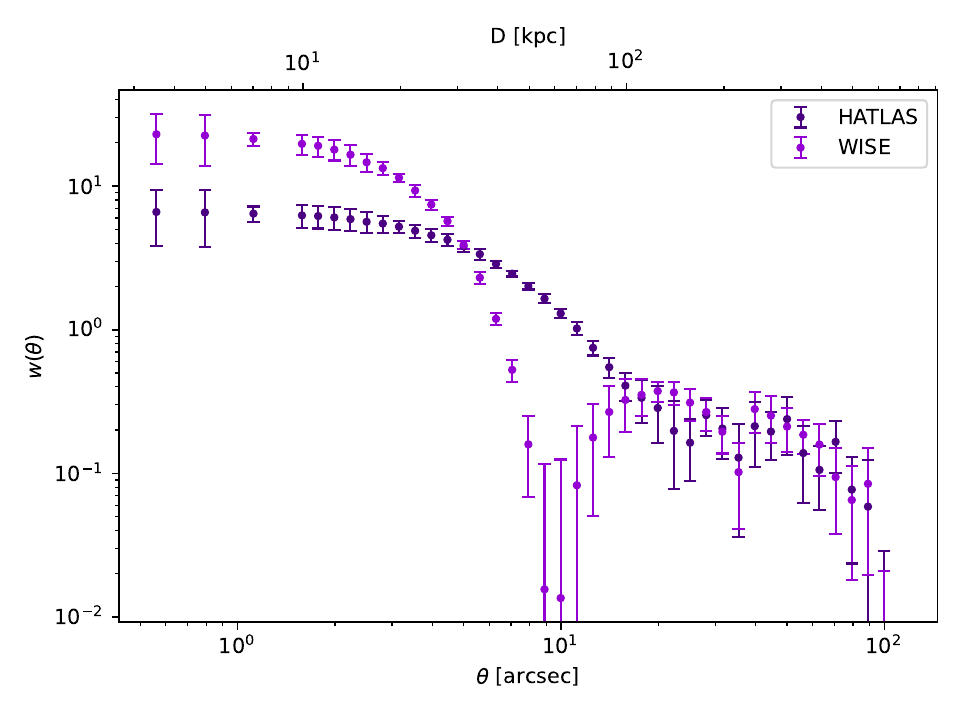}
 \includegraphics[width=0.4\textwidth]{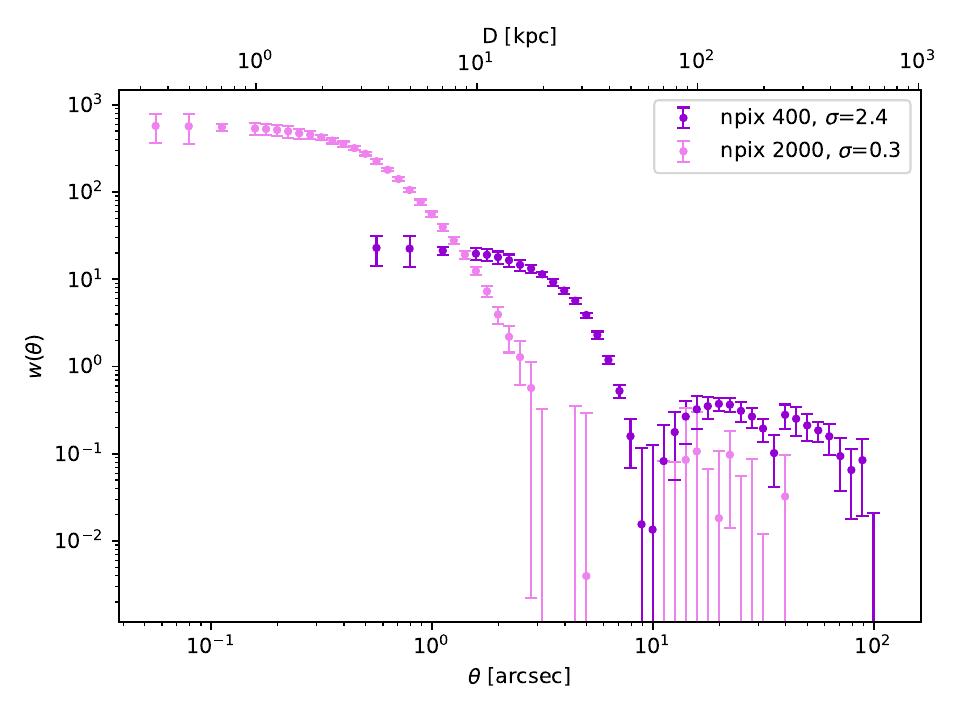}
 \caption{Cross-correlation computed for the different type of lenses. From top to bottom: galaxies (red shades), QSOs (blue shades), WEN clusters (gold shades) and ZOU clusters (magenta shades). The left column shows the comparison between the data obtained using as a background sample the WISE (lighter shade) and H-ATLAS (darker shade) positions, with npix=400, pixel size 0.5 arcsec, and $\sigma=2.4$ arcsec. In the right column we use the WISE SMG selection for the background sample and it shows the comparison between the case adopting npix=2000, pixel size 0.05 arcsec, and $\sigma=0.3$ arcsec (lighter shade) and that using npix=400, pixel size 0.5 arcsec, and $\sigma=2.4$ arcsec (darker shade).}
 \label{fig:measurements}
\end{figure*}

The CCFs measured for the four different lens catalogues are presented in Fig. \ref{fig:measurements}. From top to bottom, the measurements correspond to galaxies (red shades), QSOs (blue shades), WEN clusters (gold shades), and ZOU clusters (magenta shades). The measurements using the H-ATLAS positions and the new WISE ones for the same resolution (npix=400, pixsize=0.5 arsec and 2.4 arcsec as the positional uncertainty) are compared in the left panels. The measurements for the different samples using the WISE positions for the background sample are depicted in the right panels: the low-resolution measurements are shown with the darkest colour and the higher resolution ones (npix=2000, pixsize=0.05 arsec and 0.3 arcsec as the positional uncertainty) are shown with the lightest colour.

In the left panel of Fig. \ref{fig:measurements}, the effect of improved positional accuracy is demonstrated while maintaining the same spatial resolution characteristics. This improvement results in a higher concentration of pairs at smaller angular separations and reveals a clear lack of signal around 10 arcsec, as waspreviously indicated by \citet{FER22} and \citet{CRE22} in their respective analyses. In fact, the measurements using the H-ATLAS catalogue as a background sample for the galaxies, QSO, and the WEN cluster foreground sample are equivalent to the ones analysed in \citet{CRE22}. As expected, at larger angular separations, measurements obtained using either the H-ATLAS or WISE background sources yield practically the same results. Furthermore, there is a hint of a secondary lack of signal at larger angular separations, which varies depending on the lens sample (for example, in the WEN cluster measurements at around 60 arcsec).

However, the positional uncertainty of the WISE catalogue is superior to that of the Herschel SPIRE catalogue. Thus, by adopting WISE positions for our background sources, we can increase the angular resolution of our measurements. The right panel of Fig. \ref{fig:measurements} compares the measurements using the same resolution set-up as before (npix = 400, pixsize = 0.5 arcsec, and a positional uncertainty of 2.4 arcsec) with the new measurements that improve the resolution (npix = 2000, pixsize = 0.05 arcsec, and a positional uncertainty of 0.3 arcsec). In the later case, the pixel size is only a few hundred parsecs (the top scale is given in physical units considering the mean redshift of each sample), although the WISE positional uncertainty implies a real resolution of a few kpc. This level of resolution for studying mass density profiles is impressive for measurements based on gravitational lensing.

Based on these comparisons, two main conclusions can be drawn across all cases. Firstly, at larger angular separations the signal becomes very weak due to a lack of statistical data at such high resolutions. In any case, the measurements are comparable with the case of lower resolution. Therefore, for detailed analysis above tens of arcsec (which is beyond the scope of this paper), it is more effective to reduce the resolution to derive more robust results.

Secondly, there is a clear shift of the central excess towards smaller angular separations and higher values. This leads to an unexpected enlargement of the region lacking a signal, which now becomes evident around 3-5 arcsec in all cases but still ends at similar angular separations to those observed in the lower-resolution case (10-20 arcsec). Moreover, while the shape of the central excess in the lower-resolution case is predominantly influenced by the smoothing effect of positional uncertainty (approximately four times 2.4 arcsec, which is around 10 arcsec), we can now extract additional information as the measurements extend beyond 1.2 arcsec (four times 0.3 arcsec). In fact, there are clear differences between the various lens samples, which we discuss further in the next subsection.

The ZOU cluster catalogue provides two different positions to be considered as the centre of the cluster: the position of the brightest cluster galaxy (BCG) and the position of the maxima in the local mass distribution. To assess the effect of varying the central position of the lens in our stacking methodology, we investigated both options (see Fig. A1). The results show that at larger scales, both measurements are equivalent. However, the signal is slightly stronger for the BCG positions, indicating that, from a strong lensing perspective (as expected at these angular separations), the position of the central galaxy is more relevant than the centre of mass. Furthermore, both measurements agree regarding the signal-free region between 3-10 arcsec, confirming its presence.

\subsection{Discussion}

\begin{figure*}[htp]
 \centering
 \includegraphics[width=0.49\textwidth]{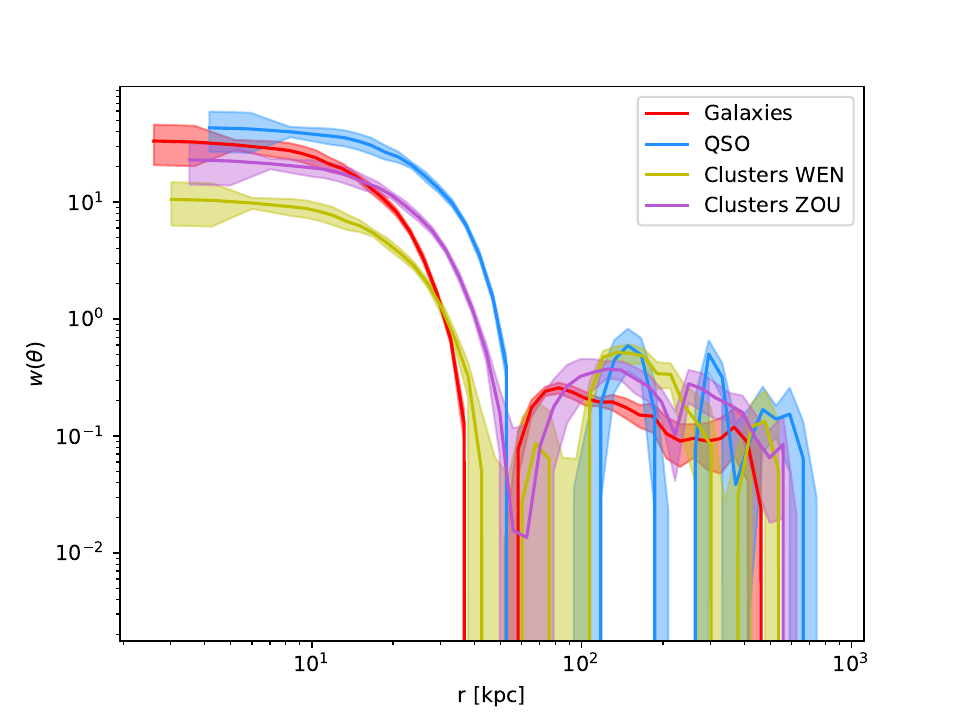}
 \includegraphics[width=0.49\textwidth]{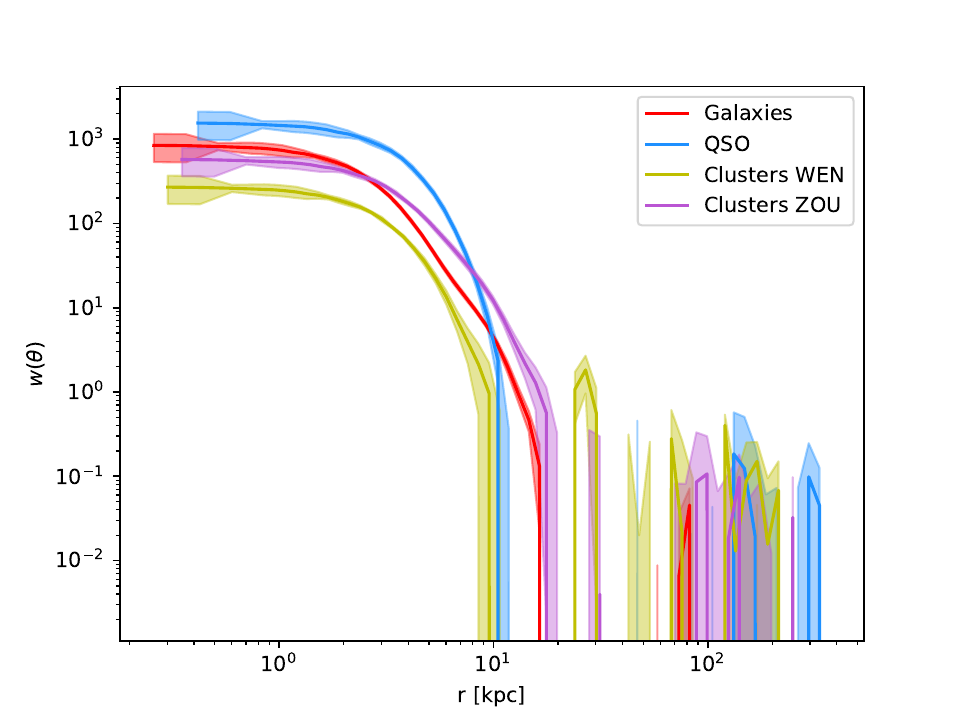}
 \caption{Comparison between the stacking obtained with the different lens samples and the background sample obtained with WISE. The distance is in kpc. Left panel: The case with npix=400, pixel size 0.5 arcsec, and $\sigma$=2.4 arcsec. Right panel: The case with npix=2000, pixel size 0.05 arcsec, and $\sigma$=0.3 arcsec.
}
 \label{fig:comp4}
\end{figure*}

The stacking results presented in this study describe the average characteristics of the lenses and should be considered as a plausible average scenario, rather than as applying to individual galaxies. Figure \ref{fig:comp4} provides a comparison of the cross-correlation measurements using the WISE positions for the background sample across different lens samples, presented in physical units considering the mean redshift of each sample. The left panel shows the low-resolution results, while the right panel shows the high-resolution ones. The colour code of the samples remains consistent with the previous figure and throughout the rest of the paper.

While the new results are qualitatively similar to those presented in \citet{CRE22}, the improved methodology and better positional accuracy of the background sample necessitate a review of the comparison between the different profiles. Similar to \citet{CRE22}, two distinct regimes are observed: a central excess and an outer, more traditional power-law profile, separated by a region with a lack of signal. Furthermore, the strength of the central signal varies among the lens samples, with QSOs producing the strongest signal and WEN clusters exhibiting weaker signals.

For the galaxy lens sample, the outer signal nearly disappears in the high-resolution case and is lower than the other lens samples in the low-resolution one. However, the central signal is not completely point-like and extends beyond the smoothing effect. Approximately 1.2 arcsec (roughly 5 kpc) corresponds to four times the positional uncertainty of 0.3 arcsec. A preliminary interpretation suggests that galaxies can primarily be considered isolated galactic haloes, with a few very close satellites present.

The QSO sample maintains a point-like central shape even in the high-resolution case, indicating compactness of the lenses. Additionally, it exhibits the most pronounced lack of signal, ranging between 10-100 kpc. Beyond 100 kpc, the signal is similar to that of the cluster lens samples. Therefore, considering that QSOs are AGNs, a plausible interpretation is that QSOs reside in overdensity environments, and the central galaxy recently underwent an interaction with a surrounding satellite galaxy, initiating the AGN episode. The signal-free region could be a result of the absence of close satellites around the central main galaxy (already cannibalised) or, equivalently, the interaction with a satellite triggering the AGN episode, which decreases the surrounding surface mass density and affects the lensing probability.

Regarding the two galaxy cluster samples, they exhibit a similar behaviour above 100 kpc, as expected. However, at smaller physical distances, their behaviour differs. According to the analysis of Fernandez et al. (2021), the stronger signal at the centre suggests a higher mean richness of the ZOU sample compared to the WEN sample, as is confirmed by the information in the catalogues: the WEN cluster catalogue mean richness is 14.44 while the ZOU cluster catalogue mean richness is 22.99. At high resolutions, the ZOU sample shows an extended profile beyond the point-like one. On the other hand, the WEN sample displays a secondary maximum at approximately 70 kpc, which splits into 2-3 maxima at higher resolutions, ranging from 30 to 80 kpc. This secondary signal is also hinted at in the ZOU sample measurements. Considering the mean redshift difference between the two samples, these results could indicate an evolutionary process, with WEN clusters being more relaxed and stable compared to ZOU clusters, which show more near satellites to the BCGs or whose closet satellites have less stable orbits.


\begin{figure}[htp]
  \centering
\includegraphics[width=\linewidth]{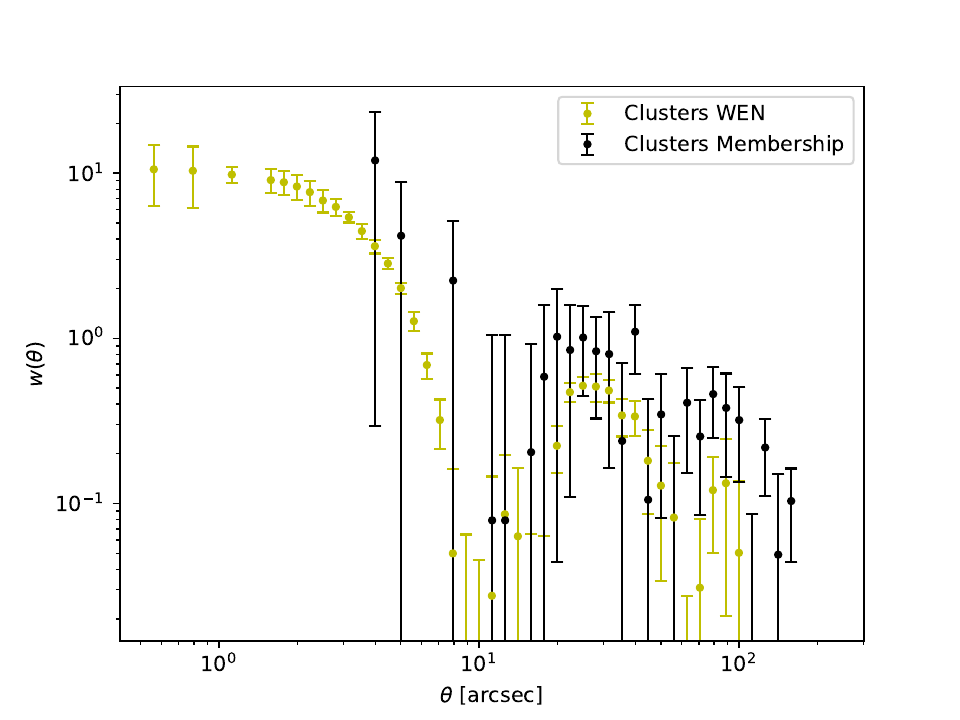}
  \caption{Stacking of the satellites in the Membership catalogue (in black) using WEN, compared with the WEN catalogue (in gold) with npix 400, pixsize 0.5, and $\sigma=2.4$.}
  \label{fig: sats}
\end{figure}

Finally, all lens samples confirm a lack of signal starting between 10-20 kpc and ending at 60-80 kpc, corresponding approximately to 10 arcsec at low resolutions for all the samples. This observation strongly suggests that it cannot be attributed to a systematic effect produced by any particular lens sample. This finding was previously anticipated in \cite{FER22} and \cite{CRE22}. Additionally, after a thorough series of tests, we ruled out the possibility that this absence of signal was produced by our methodology itself. Therefore, we began exploring other potentially related observations that could confirm the physical significance of our measurements.

Since cross-correlation measurements directly track the underlying total mass density distribution, a lack of signal in the former would imply an under-density in the latter. Consequently, the observed lack of signal with lensing should also be observable in the baryonic mass distribution, that is, in the galactic satellites.

Our first and most basic idea was to visually inspect images of galaxy clusters observed by the DESI Legacy Imaging Surveys, specifically in data release 8\footnote{https://www.legacysurvey.org/dr8/gallery/}. The inspection confirmed that most of the BCGs can be considered isolated galaxies until 5-10 arcsec from their centres. Furthermore, it revealed that the number of close satellites within 15 arcsec increases with the richness of the cluster, potentially explaining the difference between the WEN and ZOU samples. This visual inspection unexpectedly confirmed the presence of an empty ring region devoid of satellites around the BCGs, which could indicate a mass under-density and, consequently, a connection with the observed lack of lensing signal.

\citet{WAN14} study the radial distribution of satellites around bright isolated galaxies, providing a more quantitative results. However, their derived radial distribution measurements resolution (with a lowest physical scale of 10-20 kpc and just two to three data points below 100 kpc) is very interesting but not useful enough to be compared with our lensing results. Therefore, we decided to perform our own test by stacking the positions of satellites around the BCGs, maintaining the same characteristics as in our lensing measurements.

This preliminary test relied on observed galaxy cluster satellite positions rather than any lensing effect or inference from a lensing analysis. The study identified 28 clusters with information about the cluster members and their positions, which were used to create a single stacking map with the observed satellite positions. The positions of cluster members relative to the cluster centre, the BCG position, were considered for this test to observe the radial distribution of member projections. The clusters involved in this study are Berkeley 67, King 2, NGC 2420, NGC 2477, NGC 2682, NGC 6940 \citep{JAD21}, IC 2391 \citep{PLA07}, NGC 3532 \cite{FRI19}, NGC 6366 \citep{SAR15}, NGC 6530 \citep{ZHA06}, BPMG, Cha I, IC 2395, IC 348, IC 4665, LCC, NGC 1333, NGC 1960, NGC 2232, NGC 2244, NGC 2362, NGC 2547, Pleiades, THA, Taurus, UCL, Upper Sco \citep{MEN17}, and NGC 2548 \citep{WU02}. Only cluster members with a probability greater than 70\% of belonging to their respective clusters were included in the analysis.

The process of generating the stacking map was akin to the lensing analysis, but this time, it considered the observed satellite positions around the corresponding BCGs. We conducted this test using a low-resolution set-up (with a pixel size of 0.5 arcsec, 400 pixels, and a smoothing kernel of 2.4 arcsec). Galaxy satellites are baryonic bound structures, and their distribution within a galaxy cluster halo is directly influenced by the mass density profile, primarily at the first order. Therefore, examining the distribution of these satellites can provide valuable insights into the underlying mass distribution of the cluster.

This approach also bears similarities to the one adopted by \citet{DIE23}. They conducted a study comparing the distribution of dark matter (DM) with the intracluster light (ICL) and globular clusters. Just as DM particles, the stars responsible for the ICL (along with globular clusters and galactic core remnants) can be regarded as non-interacting particles that solely respond to gravity. Consequently, one would anticipate a close correlation between the distribution of ICL (and similar baryonic tracers) and DM, as supported by previous studies \citep{MON19,AA20}.

The radial analysis of the satellite stacking map now has a similar resolution and the same number of data points as the lensing measurements. Therefore, we directly compared them with the low-resolution lensing measurements for the WEN galaxy cluster sample, as is shown in Fig. \ref{fig: sats} (black points). Despite the small sample size of galaxy clusters, it is remarkable that a relatively clear measurement of the radial distribution was obtained. Moreover, the results from the satellites closely resembled the lensing ones, exhibiting similar behaviour in the outer part (above 20 arcsec) and hinting at a central excess. Importantly, there is also a tentative lack of signal around 10 arcsec, potentially confirming the previous finding from the lensing analysis. This resemblance further supports the notion that the observed distribution of galaxy satellites tends to mirror the mass density profile of the clusters derived from the lensing analysis.

On the one hand, it is traditionally accepted that satellite radial profiles parallel those of the DM, as is seen in the outer part \citep[see e.g.,]{WAN14}. In this case, the lack of signal found around 10 arcsec for the satellite radial profile could indicate poorer statistics due to the increasingly difficulty-to-identify satellites very close to their central galaxies. On the other hand, deviations from the traditional DM profile behaviour are also expected in the inner regions due to strong environmental effects, even more for the most massive central galaxies, as is demonstrated by \citet{WAN14}. Therefore, the lack of signal found around 10 arcsec could also be a consequence of such effects. In fact, it is interesting that in \citet{WAN14} the measurement at the best resolution, $\sim10$ kpc, is clearly lower than the previous ones at higher physical radial distances.

These checks provide confidence in the significance of the lack of signal observed in the lensing analysis, motivating further, more detailed analysis in subsequent sections. A detailed measurement of the stacked satellite radial distribution and its analysis goes beyond the scope of the current paper and will be addressed in a future work.

\begin{figure*}[htp]
  \centering
    \includegraphics[width=0.4\linewidth]{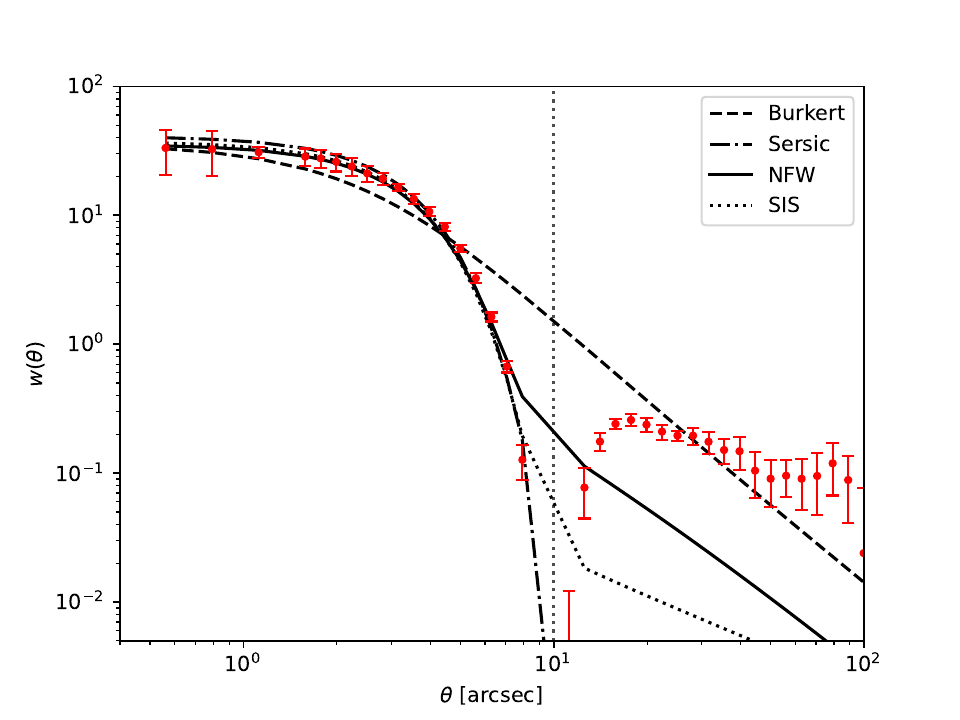}
    \includegraphics[width=0.4\linewidth]{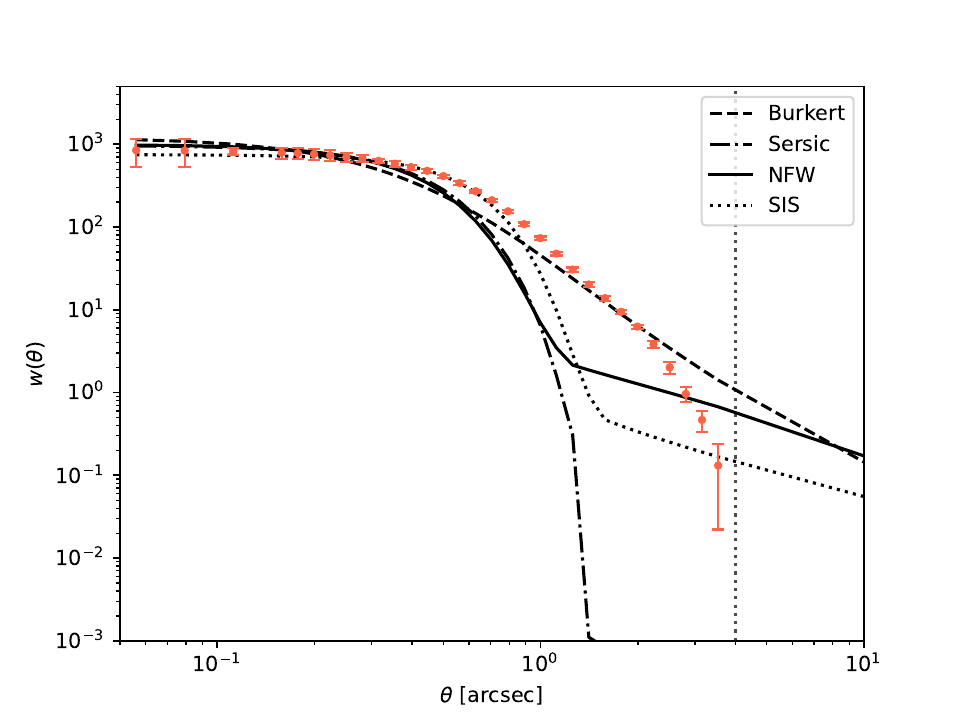}
    \includegraphics[width=0.4\linewidth]{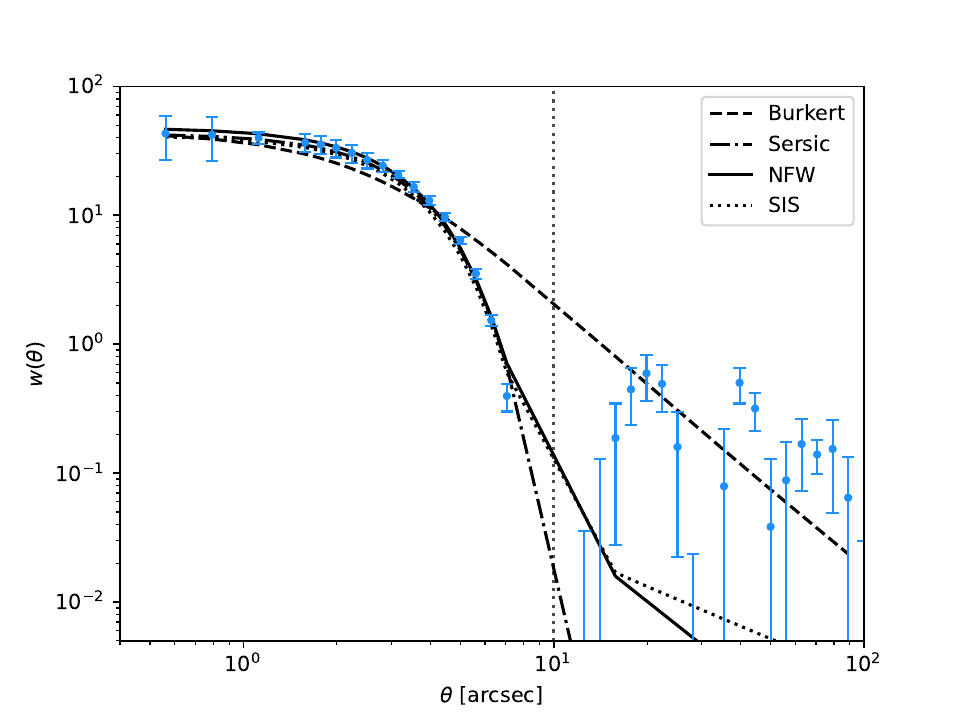}
    \includegraphics[width=0.4\linewidth]{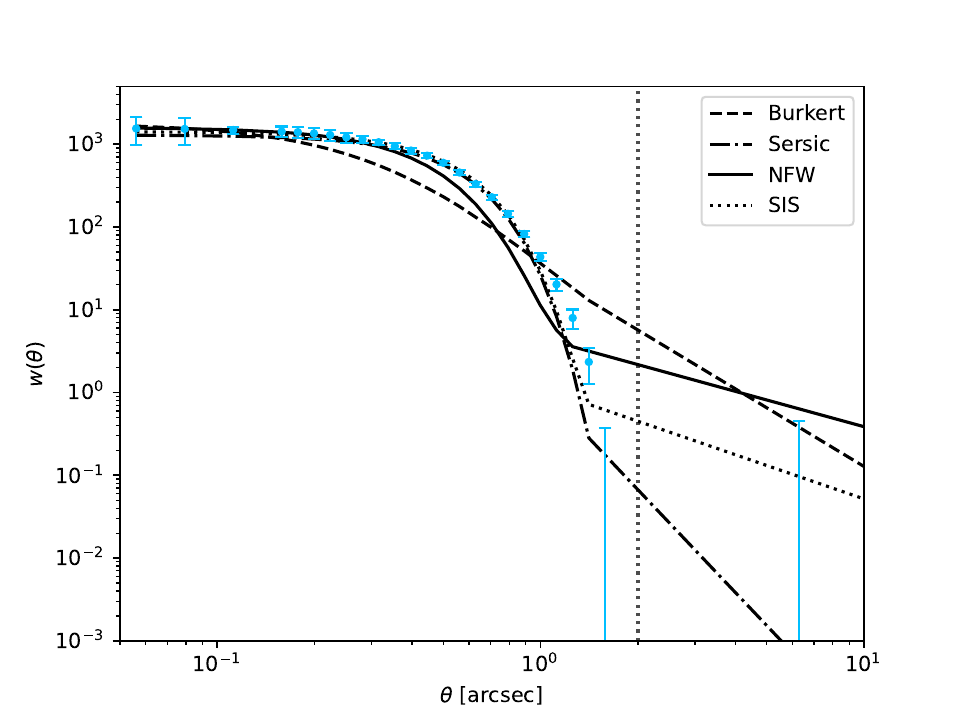}
    \includegraphics[width=0.4\linewidth]{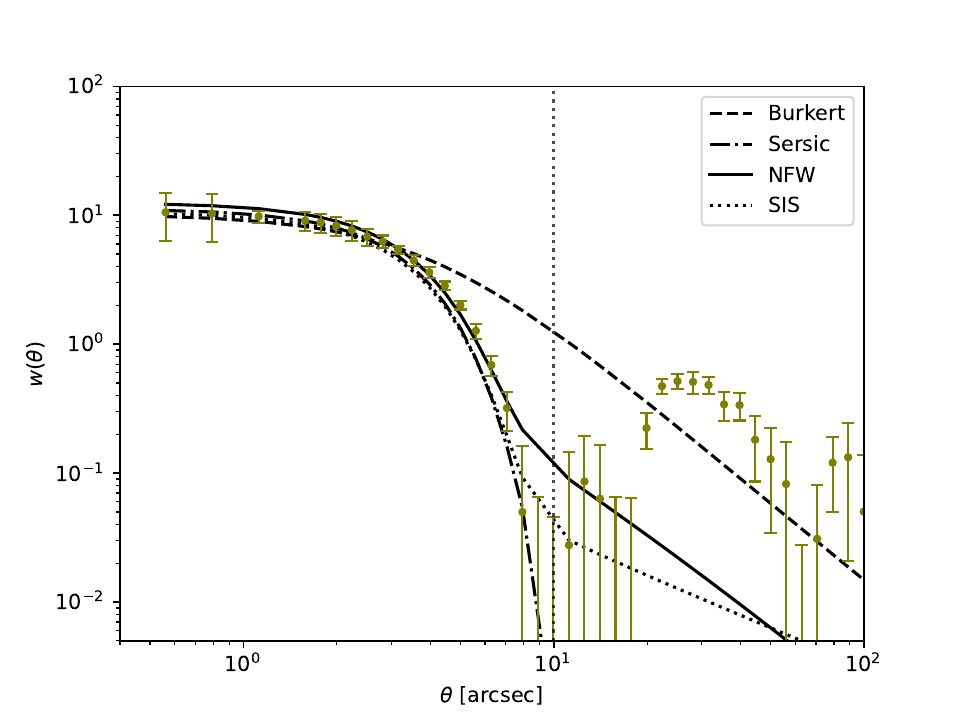}
    \includegraphics[width=0.4\linewidth]{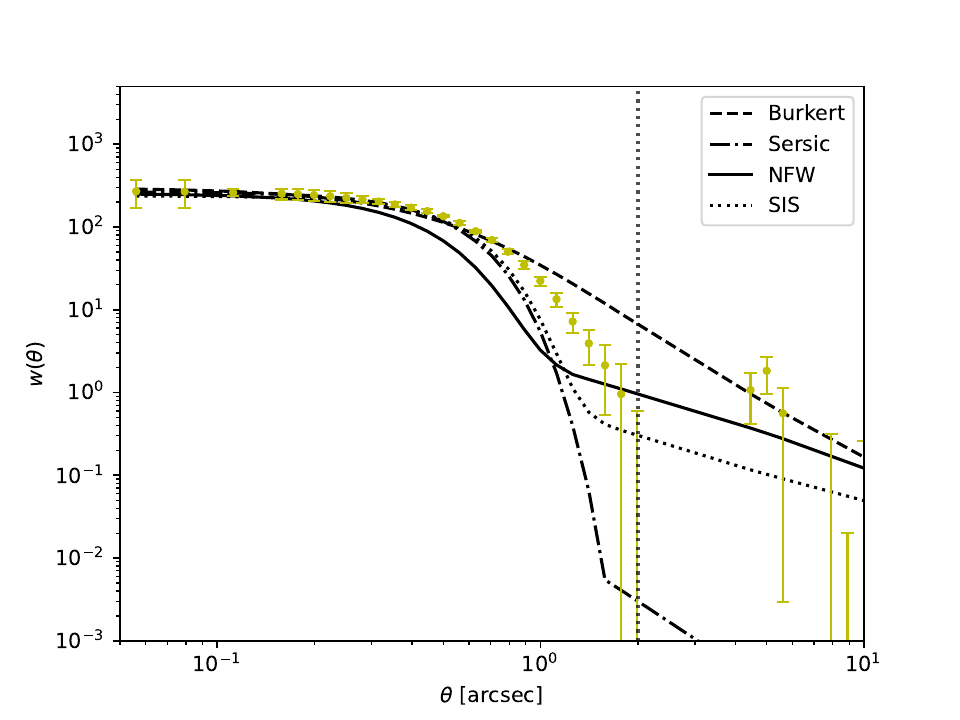}
    \includegraphics[width=0.4\linewidth]{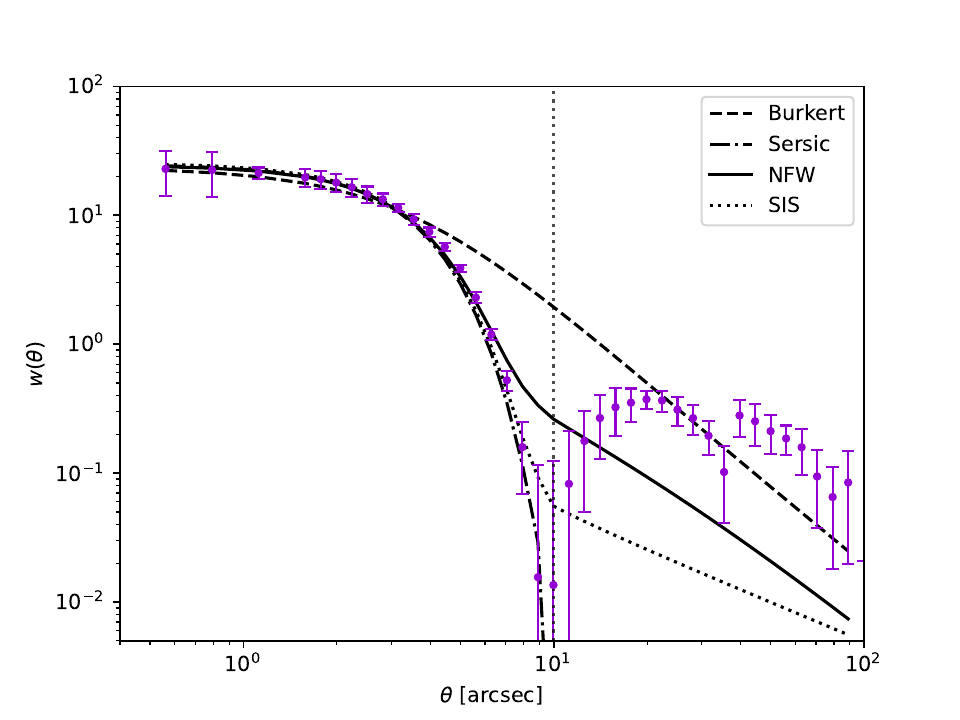}
    \includegraphics[width=0.4\linewidth]{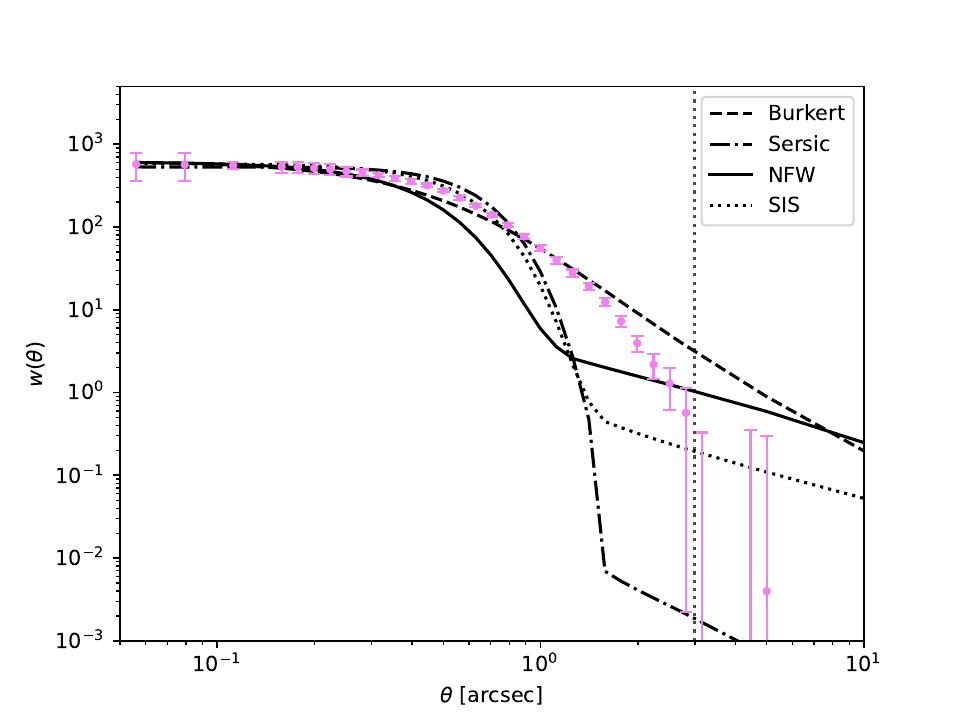}
  \caption{Mass density profile analysis for the different type of lenses. From top to bottom: galaxies (red shades), QSOs (blue shades), WEN clusters (gold shades) and ZOU clusters (magenta shades). Left panels: The npix=400 and $\sigma=2.4$ arcsec case. Right panels: The npix=2000 and $\sigma=0.3$ arcsec case. The best fit of the four profiles is shown with the black lines: Burkert, dashed line; Sérsic, dash-dotted line; NFW, solid line; and SIS, dotted line. The vertical dotted gray lines represent the range within which the points have been considered for the fitting. }
  \label{fig:prof_analysis}
\end{figure*}

\begin{table*}[]
\begin{tabular}{cc cc c cc cc }
\hline
\hline
\multicolumn{2}{c}{\multirow{2}{*}{}}                                                                                                                                   & \multicolumn{2}{c}{NFW}                          & SIS                  & \multicolumn{2}{c}{\begin{tabular}[c]{@{}c@{}}Sérsic n=4\end{tabular}} & \multicolumn{2}{c}{Burkert}                                                                     \\ \hline
\multicolumn{2}{c}{}                                                                                                                                                    & \multicolumn{1}{c}{$log_{10}(M_{r200})$} & C     & $log_{10}(M_{r200})$ & \multicolumn{1}{c}{$log_{10}(\Sigma_{e})$}        & $\theta_{e}$ (arcsec)       & \multicolumn{1}{c}{$\rho_0 \left( 10^{-2}M_\odot/\text{pc}^3\right)$} & $r_c$ (kpc) \\ \hline
\multicolumn{1}{c}{\multirow{2}{*}{Galaxies}}                                                        & \begin{tabular}[c]{@{}c@{}}npix=2000\\ $\sigma=0.3$\end{tabular} & \multicolumn{1}{c}{12.6}                & 8.0  & 12.0                & \multicolumn{1}{c}{12.0}                          & 1.0            & \multicolumn{1}{c}{11.5}                                                      & 10      \\ \cline{2-9} 
\multicolumn{1}{c}{}                                                                                 & \begin{tabular}[c]{@{}c@{}}npix=400\\ $\sigma=2.4$\end{tabular}  & \multicolumn{1}{c}{12.5}                & 8.5  & 11.4                & \multicolumn{1}{c}{11.6}                        & 1.6               & \multicolumn{1}{c}{2.1}                                                       & 40       \\ \hline
\multicolumn{1}{c}{\multirow{2}{*}{QSOs}}                                                            & \begin{tabular}[c]{@{}c@{}}npix=2000\\ $\sigma=0.3$\end{tabular} & \multicolumn{1}{c}{12.4}                & 8.9  & 12.0                & \multicolumn{1}{c}{12.0}                          & 2.1             & \multicolumn{1}{c}{5.5}                                                       & 20        \\ \cline{2-9} 
\multicolumn{1}{c}{}                                                                                 & \begin{tabular}[c]{@{}c@{}}npix=400\\ $\sigma=2.4$\end{tabular}  & \multicolumn{1}{c}{12.1}                & 10.0 & 11.7                & \multicolumn{1}{c}{11.9}                          & 5.2             & \multicolumn{1}{c}{0.8}                                                       & 100        \\ \hline
\multicolumn{1}{c}{\multirow{2}{*}{\begin{tabular}[c]{@{}c@{}}Clusters \\ Wen catalogue\end{tabular}}} & \begin{tabular}[c]{@{}c@{}}npix=2000\\ $\sigma=0.3$\end{tabular} & \multicolumn{1}{c}{12.5}                & 8.0     & 12.0               & \multicolumn{1}{c}{12.8}                          & 1.0           & \multicolumn{1}{c}{6.7}                                                       & 15       \\ \cline{2-9} 
\multicolumn{1}{c}{}                                                                                 & \begin{tabular}[c]{@{}c@{}}npix=400\\ $\sigma=2.4$\end{tabular}  & \multicolumn{1}{c}{12.4}                & 8.5  & 11.7               & \multicolumn{1}{c}{12.0}                         & 3.3             & \multicolumn{1}{c}{1.0}                                                       & 60        \\ \hline
\multicolumn{1}{c}{\multirow{2}{*}{\begin{tabular}[c]{@{}c@{}}Clusters \\ Zou catalogue\end{tabular}}} & \begin{tabular}[c]{@{}c@{}}npix=2000\\ $\sigma=0.3$\end{tabular} & \multicolumn{1}{c}{13.0}                & 5.4  & 12.0             & \multicolumn{1}{c}{11.9}                        & 3.3             & \multicolumn{1}{c}{5.0}                                                       & 20        \\ \cline{2-9} 
\multicolumn{1}{c}{}                                                                                 & \begin{tabular}[c]{@{}c@{}}npix=400\\ $\sigma=2.4$\end{tabular}  & \multicolumn{1}{c}{12.9}                & 5.4  & 12.0              & \multicolumn{1}{c}{11.9}                      & 3.3             & \multicolumn{1}{c}{0.8}                                                      & 80        \\ \hline
\end{tabular}

\caption{List of the best-fit results. From left to right: sample name, pixel number and smoothing, $\sigma$ (in arcsec), parameter values for the NFW, SIS (both masses in $M_\odot$), and Sérsic ($\Sigma_{e}$ in $\text{M}_{\odot}/\text{Mpc}^2$) and Burkert best fits. The uncertainty associated with each value is represented by the decimal place.}
\label{tab:lenses_summary}
\end{table*}

\section{Mass density profile analysis}
\label{sec:discuss}

In this section, we focus on the analysis of innermost part of the mass density profiles, specifically below $\sim10$ arcsec or $\sim50$ kpc, thanks to improvements in the resolution of the new data. We note that measurements at larger angular scales have already been comprehensively examined in a prior study by \citet{CRE22}, and so we do not duplicate this analysis here.

To analyse the cross-correlation measurements presented in the previous section, we made use of four different well-known mass density profiles. The theoretical background of the model and each individual mass density profile are summarised in Appendix \ref{ANX:profiles}. 

The Navarro-Frenk-White \citep[NFW,][]{NAV96} and the singular isothermal sphere (SIS) profiles were chosen because they are the most commonly used profiles for gravitational lensing and as a comparison with the results from our previous works \citep{FER22,CRE22}. Additionally, both works conclude that a second independent profile is needed to describe the most central region of the haloes.

As the inner part was also expected to be dominated by the baryonic matter \citep[e.g., ][]{SAL00,SOF13,CAU20,BOS21}, we wanted to test if a mass density profile without DM could describe the central region measurements. For this reason, we included the Sérsic profile, which is commonly used to describe the luminous profiles of galaxies \citep[see, for example, ][]{GRA05,TER05,COE10}. After some preliminary tests with n as a free parameter, we decided to fix it to the traditional value for elliptical galaxies, n=4, due to the lack of relevant improvements in the fits.

Furthermore, to address the typical divergence of the NFW and SIS profiles at the centre, we introduced a cored profile and selected the Burkert model as the most representative choice. It provides a more stable and well-behaved description of the central mass distribution, making it useful in cases where other profiles may encounter difficulties in fitting or explaining observational data in the inner regions of galaxies \citep[see, for example,][]{SAL00,BOR01,GEN07,UME16,KHE23,KRU23}. It is important to note that selecting slight modifications of each profile does not affect the main conclusions drawn from the current analysis.

To estimate the best-fit parameters for each mass density profile, we smoothed the theoretical prediction by the same filtering kernel applied to the data in order to take into account the positional uncertainty in each type of measurements. 
The resulting cross-correlation functions using the best-fit mass density profile parameters are presented in Fig. \ref{fig:prof_analysis}, where the left and right columns display the low (npix=400 and $\sigma=2.4$ arcsec) and high (npix=2000 and $\sigma=0.3$ arcsec) resolution cases, respectively. The lens samples order from top to bottom, and the colour code remains consistent with Figure \ref{fig:measurements}. As is indicated by a vertical dotted line in each panel, the fits were performed using only those measurements below 10 arcsec for the low-resolution case and 5 arcsec for the high-resolution ones.
A summary of the best-fit values of all lens types and profiles is provided in Table \ref{tab:lenses_summary}. 

The unique characteristics of the measurement data posed significant challenges for conventional fitting algorithms, rendering them ineffective in providing meaningful results. Given the complexity and subtleties of the data, fitting algorithms required substantial guidance to yield even qualitative outcomes. Consequently, we opted for a manual fitting approach to obtain preliminary insights. Regrettably, due to the intricacies involved, we were unable to estimate uncertainties for the results. However, the decimal places provided in Table \ref{tab:lenses_summary} can be considered a rough first-order approximation of the uncertainties. Additionally, the limitations of the dataset meant that exploring potential correlations between parameters was not feasible at this stage. Therefore, until more robust models or comprehensive analyses can be conducted in the future, our discussion remains primarily qualitative, offering valuable preliminary insights into the mass density profiles and their characteristics.

In the case of the low-resolution data, the central behaviour of the mass density distribution appears to be “point-like” for all types of lenses, meaning that the profiles simply adopt the shape of the applied smoothing kernel. As a result, all profiles, except for the Burkert profile, provide a good fit to the data below 10 arcsec. However, it is evident from the results that none of these profiles can simultaneously explain the larger-scale data. This conclusion aligns with the findings of \citet{FER22} and \citet{CRE22}. The Burkert profile, with its distinct shape, provides a decent global fit at both small and large angular separations considered here, but it overestimates the range around 10 arcsec where the lack of signal is observed.

With the high-resolution data, the quality of the fits deteriorates even at smaller angular separations. As was mentioned earlier, only the QSOs maintain a “point-like” behaviour, with the Sérsic and SIS profiles still providing a good fit to the data. At these angular scales, the differences between the various profiles are mainly due to the characteristic angular separation of the caustic(s), which varies for each profile. However, these caustics are not observable in the figures due to the smoothing step caused by positional uncertainty. For galaxies and clusters, the theoretical profiles fail to explain the extended emission beyond one arcsec. The Burkert and NFW profiles can be clearly discarded due to their poor performance. The SIS and Sérsic profiles yield very similar results, but the SIS profile overestimates the larger angular separation scales above 2-3 arcsec. 

Given the inherent challenges in obtaining reasonable fits to the measurements using the selected mass density profiles, quantitative approaches such as Bayesian evidence or similar statistical techniques were not employed to compare and select the best-fitting model. These methods typically rely on well-behaved datasets and robust model-data agreement, which, in this context, were not attainable due to the complexities of the measurement data.

The inability of any of the profiles to explain the extended emission suggests that it is not directly related to the compactness of the lenses. One possible interpretation regarding the nature of this additional emission could be the contribution of close satellites along the line of sight. Since haloes are three-dimensional objects, the projection of some satellites can appear near the centre without any ongoing interaction. Therefore, the lack of extended emission in the case of QSOs could indicate a scarcity of such close satellites, either due to a low number or because the nearby satellites are interacting with the central galaxy, triggering the AGN phase. We also explored the possibility of introducing a second, independent profile to improve the fit of the extended emission. The shape of this extended emission cannot be explained by the power-law nature of the theoretical profiles; it arises as a direct consequence of the smoothing effect on the caustic(s). To produce a departure from the power-law behaviour at 2-3 arcseconds, all theoretical profiles would require nonphysical values for the central mass, exceeding the mass of an entire galaxy cluster in some cases. Furthermore, these masses would significantly overshoot the data in the central region. Therefore, we conclude that the extended emission cannot be modelled as an independent component described by any of our theoretical profiles.

Focusing on the best-fit values shown in Table \ref{tab:lenses_summary}, a comparison of the derived masses for the NFW and SIS profiles with previous studies reveals several interesting findings. Firstly, the masses are similar across all lens types, as they correspond to the single main central galaxy in each case. This result implies that only the most massive galaxies, with masses of the same order as the BCGs in low-richness galaxy clusters (the most abundant), are acting as lenses. Secondly, the ZOU cluster masses are slightly higher than those of the WEN sample, as well as the galaxy and QSO samples. This difference could be attributed to the smaller concentration in the ZOU clusters, although it may also be linked to the higher richness of the ZOU cluster catalogue, indicating a more massive BCG. There is a non-negligible correlation between mass and concentration, with higher masses corresponding to smaller concentrations and vice versa. Thirdly, the NFW masses are consistently higher than the SIS masses, which is in line with previous results \citep{BON19,FER22,CRE22}. 

Furthermore, the new mass estimates are generally smaller than those obtained in previous studies \citep{BON19,FER22,CRE22}. Specifically, the NFW masses are more in agreement with the SIS results from \citet{CRE22}, which were shown to underestimate the measurements with the original H-ATLAS positions. It appears that the improved positional accuracy reveals a lack of signal in certain regions, which necessitates higher concentrations and smaller masses. Notably, when attempting to fit both the inner and outer regions without accounting for the lack of signal, the results align more closely with those derived in \citet{CRE22}. Therefore, the new derived masses indicates that the central galaxies are less massive than the typical luminous red galaxies (LRGs) that usually correspond to the BCGs \citep{BLA08,CAB09,BAU14}.

The results from the other mass density profiles, namely Sérsic and Burkert, cannot be directly compared. However, the Sérsic surface density results show a consistent trend across different lens types, with values around $10^{12}$ $\text{M}_{\odot}/\text{Mpc}^2$ and a characteristic scale of 2-3 arcsec. On the other hand, the best-fit parameters for the Burkert profile exhibit greater variability with respect to lens type and data angular resolution.

Moreover, as the resolution increases for all lens samples and mass density profiles, certain patterns in the best-fit parameters emerge. For the NFW fit, there is a consistent pattern of increasing mass with decreasing concentration, which is also observed in the Sérsic and Burkert cases. In both profiles, with increasing resolution, the “mass-related” parameter ($\Sigma_{e}$ and $\rho_0$) increases while the “radius” parameter ($\theta_{e}$ and $r_c$) decreases. Regarding the SIS fit, an increase in mass is observed. Interestingly, this kind of pattern does not hold for the ZOU clusters, where the best-fit parameters remain the same between the two resolution measurements for all profiles, except for the Burkert one.

Regarding the Burkert profile, previous studies \citep{SAL00, DON09, GEN09, BUR15} have revealed a strong relationship between $\rho_0$ and $r_c$ when fitting galaxy rotation curves with this profile. This relationship is characterised by a constant value of $\mu_{0D}$, defined as $\mu_{0D} \equiv \rho_0 \cdot r_c$, which remains consistent across different galaxies. The estimated values of $\mu_{0D}$ have ranged from approximately $\mu_{0D}\approx 140 ^{+80}_{-30} \,M_{\odot}\,\text{pc}^{-2}$ \citep{DON09} to $\mu_{0D}\approx 75 ^{+85}_{-45} \,M_{\odot}\,\text{pc}^{-2}$ \citep{BUR15}. This constant relationship implies that the central surface density of DM in galaxies is largely independent of their mass \citep{DON09}.

Using the data from Table \ref{tab:lenses_summary}, we find that this constant relationship holds, but with higher values than expected from previous works. Specifically, we estimate $\mu_{0D}\approx 870 ^{+200}_{-160} M_{\odot}\,\text{pc}^{-2}$. The Burkert density profile struggles to adequately fit the observed data, indicating potential variations in both $\rho_0$ and $r_c$ with respect to previous studies. In fact, if we fix $\mu_{0D}\sim 100 M_{\odot}\,\text{pc}^{-2}$, in agreement with those results, we are able to produce a reasonable fit only for the larger angular scales in the low-resolution case. In this case, the Burkert profile overestimates the cross-correlation around 10 arsec and is well below the data at smaller angular separations. This result is equivalent to the conclusion from \citet{CRE22}, that two different mass density profile are needed to explain the inner (below 10 arcsec in the low-resolution case) and outer (above 10 arcsec in the low low-resolution case) parts of the measurements.

Finally, it is intriguing that all datasets exhibit a lack of signal around 10 arcsec in the low-resolution case, which cannot be explained by any of the considered mass density profiles. To investigate the possibility of systematic issues in the data handling that may have caused these lower cross-correlation measurements, we developed a magnification bias simulation tool. This tool applies a mass density profile to simulate lensing and computes the cross-correlation signal using the same procedure as the actual data (González-Nuevo et al., in prep). Although there are still observational systematics and improvements to be incorporated, the current version of the magnification bias simulator tool offers the basic functionality for a preliminary check.

\begin{figure}[htp]
  \centering
    \includegraphics[width=\linewidth]{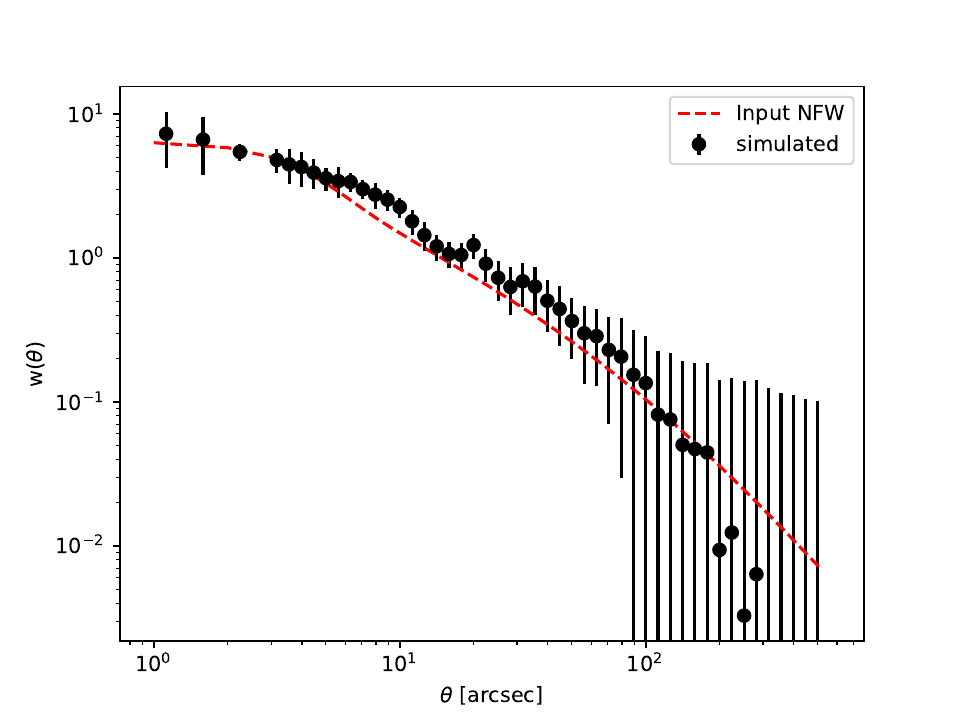}
  \caption{Cross-correlation data (black dots) simulated with 50000 lenses with of $10^{14}M_\odot$ total mass. The redshift for all the lenses is $z=0.3$ while it is $z=2.2$ for all the background sources. The simulations adopt an NFW profile (dashed red line) with a smoothing of $\sigma=2.4$ arcsec and a pixel size of 0.5 arcsec.}
  \label{fig:xcsim}
\end{figure}

In this initial test, all lenses had the same properties, with an average mass of $10^{14}M_{\odot}$, following an NFW mass density profile, and a redshift of $z=0.3$. The background sources were simulated within a radius of 100 arcsec (npix=400 with pixsize=0.5, as in the low-resolution case) using the \texttt{CORRSKY} software \citep{GN05}, incorporating the source number counts from \citealt{CAI13} and the angular power spectrum obtained from \citealt{LAP11}. Initially, we simulated background sources with flux densities $S>1$ mJy at $250\mu m$, but we only analysed those whose flux density, after the amplification factor, exceeded $\sim29$ mJy, the H-ATLAS flux limit. All background sources were placed at redshift 2.2. To mimic the SPIRE beam effect, we removed any faint background source located within 17 arcsec of a brighter one and a positional uncertainty of $\sigma=2.4$ arcsec was also considered. Figure \ref{fig:xcsim} shows the simulated data obtained through this procedure as black dots, while the input theoretical signal is depicted by the dashed red line. There is excellent agreement between the theoretical signal and the simulated data, with no indication of a lack of signal at 10 arcsec or any other angular separation. Hence, we can preliminarily conclude that no systematic issues were introduced during the computation of the cross-correlation data points.

This lack of signal poses a challenge that requires further investigation to understand its origin. It is possible that actual physical processes, not accounted for in the theoretical model, are responsible. There are interesting results that may be relevant to this type of feature. \citet{CHA22} developed a physically motivated criterion to define the edges of galaxies. They defined the edge of a galaxy as the outermost radial location associated with a significant drop in either past or ongoing in situ star formation, based on the gas density threshold required for star formation. For elliptical galaxies with stellar masses between $M_\star=10^{10-12} M_\odot$, they found typical edge radii in the range of 10-100 kpc, which aligns with our results for the high-resolution case (see Fig. \ref{fig:comp4}). 

In \citet{DIE23}, a free-form model of the SMACS0723 galaxy cluster is presented. This model avoids strong assumptions about the distribution of mass (mostly DM) in the cluster and is used to study the possible correlation between DM and the ICL, as well as the distribution of globular clusters. Figure 6 in their paper illustrates a one-dimensional scan of the light distribution versus the modelled DM distribution. Around -200 kpc, the smoothed DM model overestimates the observed ``cavity'' in the luminous matter tracers, similar to the issue encountered in our analysis. Additionally, our measurements resemble those predicted by the fuzzy or wavy DM model with a central soliton and a NFW-like external asymptotic profile \citep{SCH14, HUI21}. For the central densities derived from the Burkert profile, the estimated soliton core radius of a particle with a mass of $\sim 10^{-22} \text{eV}$ is of the order of a few kpc. Thus, the signal-free region extends to approximately three to four times this value.

\section{Conclusions}
\label{sec:concl}
This study focused on investigating the magnification bias effect in the context of gravitational lensing. Specifically, we examined the impact of magnification bias on the flux of SMGs, which exhibit steep source number counts, making them an ideal background sample for magnification bias studies. The magnification bias was estimated using the CCF between different types of foreground lenses, including QSOs, galaxies, and galaxy clusters. Employing the stacking technique, we aimed to compare the mass density profiles of these lens types and shed light on their properties. By adopting the WISE catalogue positions for background sources, we increased the angular resolution, which enhanced the concentration of pairs at smaller angular separations and revealed a clear lack of signal around 10 arcsec in the low-resolution case. The choice of the central lens position, specifically the BCG position, yielded slightly stronger signals from a strong lensing standpoint. These findings provide valuable information for understanding mass density profiles at kiloparsec scales.

The cross-correlation measurements using the WISE positions for the background sample reveal two distinct regimes: a central excess and an outer power-law profile, separated by a region with a lack of signal. The central signal strength varies among the lens samples, with QSOs producing the strongest signal and WEN clusters exhibiting weaker signals. The galaxy lens sample shows a diminishing outer signal in the high-resolution case but maintains a non-point-like central signal, suggesting a scenario of isolated galactic haloes with some very close satellites. The QSO sample remains point-like even in the high-resolution case and displays a pronounced lack of signal, possibly indicating a connection between QSOs, overdensity environments, and recent interactions with satellite galaxies, as was also concluded by \citet{MAN05}, for example. The two galaxy cluster samples exhibit similar behaviour beyond 100 kpc but differ at smaller distances, suggesting differences in mean richness and cluster stability between the WEN and ZOU samples. Furthermore, all lens samples confirm a lack of signal between 10 and 60 kpc, potentially explained by the presence of isolated central galaxies and an increasing number of close satellites with richness.

Next, we analysed the cross-correlation measurements using four different theoretical mass density profiles to extract additional information. Firstly, we selected the NFW, SIS, Sérsic, and Burkert profiles, which provide varying degrees of fit to the data. The NFW and SIS profiles, commonly used for their descriptive capabilities, offer a good fit to the central region of the haloes. The Sérsic profile, commonly used for describing luminous galaxy profiles, was included to account for baryonic matter dominance in the central region. The Burkert profile was chosen to address the divergence of previous profiles at the centre. However, none of the profiles can simultaneously explain the larger-scale data, indicating the need for additional considerations.

The analysis using high-resolution data shows a deterioration in the quality of fits, even at smaller angular separations. Only the QSOs exhibit a “point-like” behaviour, while galaxies and clusters show extended emission beyond one arcsec. The Burkert and NFW profiles perform poorly, while the SIS and Sérsic profiles yield similar results, with the SIS profile overestimating larger angular separation scales. The lack of extended emission in the case of the QSOs suggests that it may not be directly related to the compactness of the lenses; instead, it could be influenced by the presence of close satellites along the line of sight.

Comparing the derived masses for the NFW and SIS profiles with previous studies, we observe that the masses are similar across all lens types and slightly higher for the ZOU cluster sample. The NFW masses consistently exceed the SIS masses, consistent with previous findings. The new mass estimates, based on improved positional accuracy, indicate that the central galaxies are less massive than typical large red galaxies. The Sérsic surface density results show a consistent trend across different lens types, with characteristic values around $10^{12}$ $M_{\odot}/\text{Mpc}^2$ and a scale of 2-3 arcsec. The best-fit parameters for the Burkert profile exhibit greater variability with respect to lens type and data angular resolution.

Increasing the resolution reveals certain patterns in the best-fit parameters. For the NFW and Sérsic fits, there is a consistent pattern of increasing mass with decreasing concentration. Notably, the ZOU clusters show no significant variation in the best-fit parameters between the two resolution measurements, except for the Burkert profile. Regarding the Burkert profile, previous studies have shown a strong relationship between the central surface density of DM ($\rho_0$) and the core radius ($r_c$). This relationship, characterised by a constant value of $\mu_{0D}$, appears to hold in this analysis but with higher values than in previous works. Fixing $\mu_{0D}$ to previous estimates, a reasonable fit is achieved only for larger angular scales, while the profile fails to match the data at smaller separations.

Overall, the analysis of the mass density profiles highlights the limitations of the selected profiles in explaining the observed data. This result aligns with the need for two different mass density profiles to explain the inner and outer parts of the measurements, suggesting the presence of additional factors, such as close satellites and the characteristics of the environmental outer emission, that need to be considered in future investigations. The derived masses and best-fit parameters provide valuable insights into the nature of lensing systems and contribute to our understanding of the central galaxies and their mass distributions.

Finally, our analysis reveals a remarkable lack of signal around 10 arcsec that cannot be accounted for by any of the considered mass density profiles. This intriguing feature challenges our current understanding of the relationship between gravitational lensing and the distribution of mass. To confirm these findings, we conducted an additional test involving stacking the positions of satellites around the BCGs. This test utilised observed galaxy cluster satellite positions and revealed a radial distribution similar to the lensing measurements, supporting the lack of signal found in the lensing analysis. Moreover, to investigate potential systematic issues in our data handling, we developed a magnification bias simulation tool. Preliminary tests using this tool showed no indication of a lack of signal at 10 arcsec or any other angular separation, suggesting that the process of computing the cross-correlation data points is unlikely to introduce systematic errors. While the detailed analysis of the stacked satellite radial distribution or the development of a full simulator tool goes beyond the scope of this paper, these checks provide confidence in the significance of the lack of signal and pave the way for further quantitative analysis in future publications.

\begin{acknowledgements}
DC, JGN, LB, JMC acknowledge the PID2021-125630NB-I00 project funded by MCIN/AEI/10.13039/501100011033/FEDER, UE. 
LB also acknowledges the CNS2022-135748 project funded by MCIN/AEI/10.13039/501100011033 and by the EU “NextGenerationEU/PRTR”.
JMC also acknowledges financial support from the SV-PA-21-AYUD/2021/51301 project.\\

The \textit{Herschel}-ATLAS is a project with \textit{Herschel}, which is an ESA space observatory with science instruments provided by European-led Principal Investigator consortia and with important participation from NASA. The H-ATLAS web-site is http://www.h-atlas.org. GAMA is a joint European-Australasian project based around a spectroscopic campaign using the Anglo-Australian Telescope.\\

This research has made use of the python packages \texttt{ipython} \citep{ipython}, \texttt{matplotlib} \citep{matplotlib} and \texttt{Scipy} \citep{scipy}.
\end{acknowledgements}

\bibliographystyle{aa} 
\bibliography{Stack.bib} 

\appendix
\section{Theoretical framework}
\label{ANX:profiles}

The gravitational lensing effect produces a modification of the integral number counts of background sources in a flux-limited sample due to the presence of a mass distribution between these sources and the observer. This alteration occurs because of the combined effects of magnification, which enhances the flux of fainter sources, and dilution, which expands the apparent solid angle on the sky. Consequently, the number of background sources per unit solid angle and redshift, with an observed flux density greater than $S$, is modified at each two-dimensional angular position $\vec{\theta}$ on the celestial sphere as \citep{Bar01}

\begin{equation}
    n(>S,z;\vec{\theta})=\frac{1}{\mu(\vec{\theta})}n_0\Big(>\frac{S}{\mu(\vec{\theta})},z \Big),\label{eq2}
\end{equation}
where $n_0$ represents the integrated number counts in the no-magnification case and $\mu(\vec{\theta})$ denotes the magnification field at angular position $\vec{\theta}$. Assuming a redshift-independent power-law behaviour of the unlensed integrated number counts, expressed as $n_0(>S,z)=AS^{-\beta}$, equation \ref{eq2} can be rewritten as

\begin{equation}
\frac{n(>S,z;\vec{\theta})}{n_0(>S,z)}=\mu^{\beta-1}(\vec{\theta}).
\label{eq3}
\end{equation}

The relationship between the magnification field due to a sample of lenses and the cross-correlation observable that we aim to measure becomes clear when we understand the physical meaning of the above ratio. The left side of equation \ref{eq3} represents the excess or deficiency of background sources (in the direction $\vec{\theta}$ as viewed by a lens at redshift $z_l$) at a redshift $z_b$ > $z_l$, compared to the case without lensing. The angular CCF between a foreground sample of lenses (at redshift $z_l$) and a background sample of objects (at redshift $z_b$) samples is defined as
\begin{equation}
    w_x(\vec{\theta};z_l,z_b)\equiv\langle\delta n_f(\vec{\phi})\,\delta n_b(\vec{\phi}+\vec{\theta})\rangle,
\end{equation}
where $\delta n_b$ and $\delta n_f$ denote the background and foreground object number density variations, respectively. Since we are stacking the lenses at a fixed position, it follows from the above argument that

\begin{equation}
    w_x(\vec{\theta};z_l,z_b)=\mu^{\beta-1}(\vec{\theta})-1.
\end{equation}

Let us assume that a lens located at an angular diameter distance $D_d$ from the observer deflects the light rays from a source at an angular diameter distance $D_s$. If $\vec{\theta}=\vec{\xi}/D_d$ denotes the angular position of a point on the image plane, the convergence field can be defined as

\begin{equation}
    \kappa(\vec{\theta})=\frac{\Sigma(D_d\vec{\theta})}{\Sigma_{\text{cr}}},
\end{equation}
where $\Sigma(\vec{\xi})$ denotes the mass density projected onto a plane perpendicular to the light ray and $\Sigma_{\text{cr}}$ is the critical mass density characterised as
\begin{equation}
    \Sigma_{\text{cr}}=\frac{c^2}{4\pi G}\frac{D_s}{D_dD_{ds}}
\end{equation}
where $D_{ds}$ represents the angular diameter distance from the lens to the background source.\\

If we assume that a lens is axially symmetric, we can choose the origin as the symmetry centre yields $\kappa(\vec{\theta})=\kappa(\theta)$ and the magnification field, $\mu(\theta)$, is related to the convergence via \citep{Bar01}

\begin{equation}
    \mu(\theta)=\frac{1}{(1-\bar{\kappa}(\theta)(1+\bar{\kappa}(\theta)-2\kappa(\theta))}.
\end{equation}

In this context, $\bar{\kappa}(\theta)$ refers to the mean surface mass density within the angular radius $\theta$ designated as

\begin{equation}
    \bar{\kappa}(\theta)= \dfrac{2}{\theta^2} \cdot\int_{0}^{\theta} d\theta' \theta' \kappa(\theta')
\end{equation}



\subsubsection{Navarro-Frenk-White profile}

 The Navarro-Frenk-White \citep[NFW,][]{NAV96} mass density profile is one of the most used profiles in cosmology. It is a two-parameter model given by
 \begin{equation}
     \rho_{\text{NFW}}(r;r_s,\rho_s)= \frac{\rho_s}{(r/r_s)(1+r/r_s)^2},
 \end{equation}

which can be shown to satisfy \citep{SCH06}

\begin{equation}
    \kappa_{\text{NFW}}(\theta)=\frac{2r_s\rho_s}{\Sigma_{\text{cr}}}f(\theta/\theta_s)\quad\quad\quad \bar{\kappa}_{NFW}(\theta)=\frac{2r_s\rho_s}{\Sigma_{cr}}h(\theta/\theta_s),
\end{equation}
where $\theta_s\equiv r_s/D_d$ is the angular scale radius,
\begin{equation}
    f(x)\equiv \begin{cases} \frac{1}{x^2-1}-\frac{\arccos{(1/x)}}{(x^2-1)^{3/2}}\quad\quad&\text{if } x>1\\
    \,\,\frac{1}{3} \quad\quad&\text{if } x=1\\
    \frac{1}{x^2-1}+\frac{\text{arccosh}(1/x)}{(1-x^2)^{3/2}} \quad\quad&\text{if } x<1
    \end{cases}
\end{equation}
and
\begin{equation}
    h(x)\equiv \begin{cases} \frac{2}{x^2}\Big(\frac{\arccos{(1/x)}}{(x^2-1)^{1/2}}+\log{\frac{x}{2}}\Big)\quad\quad&\text{if } x>1\\
    \,{\scriptstyle 2\,(1-\log{2})} \quad\quad&\text{if } x=1\\
    \frac{2}{x^2}\Big(\frac{\text{arccosh }(1/x)}{(1-x^2)^{1/2}}+\log{\frac{x}{2}}\Big) \quad\quad&\text{if } x<1
    \end{cases}.
\end{equation}

\subsubsection{Singular isothermal sphere profile}

 The singular isothermal sphere (SIS) model is another well-known and used mass density profile due to its simplicity. It is given by
 \begin{equation}
     \rho_{\text{SIS}}(r)=\frac{\sigma_v^2}{2\pi G r^2},
 \end{equation}

 The convergence and mean surface density inside $\theta$ are easily shown to be \citep{SCH06}
 
\begin{equation}
    \kappa_{\text{SIS}}=\frac{\theta_E}{2|\theta|}\quad\quad\bar{\kappa}_{\text{SIS}}(\theta)=\frac{\theta_E}{|\theta|},
\end{equation}
where
\begin{equation}
    \theta_E=4\pi\,\bigg(\frac{\sigma_v}{c}\bigg)^2\frac{D_{ds}}{D_s}
\end{equation}
is the so-called Einstein radius of the model.

\subsubsection{The Sérsic profile}
Sérsic's $R^{1/n}$ model \citep{SER63, SER68} is a three-parameter model defined by the surface brightness profile. It has been widely used to describe the light profiles of galaxies, including both elliptical and disk galaxies \citep{CIO91,TRU02}. Its use as a mass profile is described, for example, by \citet{GRA05} or \citet{TER05} with lensing properties calculated by \citet{CAR04}, \citet{ELI07} and \citet{COE10}, among others.

The surface mass density of a Sérsic model can be expressed as:

\begin{equation}
    \Sigma(\theta)=\Sigma_{e}\cdot \text{exp} \left[ -b_n \left(\left( \frac{\theta}{\theta_{e}} \right) ^{1/n}-1\right) \right]
\end{equation}
where $\theta_e$ is the angular position of the  effective radius ($R_e$), $\Sigma_e$ the surface mass density at the effective radius and $n$ the Sérsic index. This last parameter describes the shape of the surface brightness or surface mass density profile of a galaxy, cluster, or other astronomical object \citep{SER63}. The parameter $b_n = b(n)$ is not a free parameter in the model and is defined as $\Gamma(2n)= 2\gamma(2n,b_n)$, where $\Gamma(a)$ and $\gamma(a,x)$ are the gamma function and the incomplete gamma function respectively \citep{ABR64}.\\

It can be shown that the Sérsic profile satisfies \citep{SCH06}

\begin{equation}
    \kappa(\theta)_{Sersic}=\dfrac{\Sigma_{e}}{\Sigma_{cr}}  \cdot exp \left[ -b_n \left(\left( \frac{\theta}{\theta_{e}} \right) ^{1/n}-1\right) \right]
\end{equation}
\begin{equation}
    \bar{\kappa}(\theta)_{Sersic}=\dfrac{\Sigma_{e}}{\Sigma_{cr}}  \cdot \dfrac{2n\cdot e^{b_n}}{\theta^2} \left[ \dfrac{ (2n-1)! \cdot \theta_e}{b_n^{2n}} - \theta^2\cdot\text{E}_{2n-1}\left[ b_n \left( \dfrac{\theta}{\theta_e}\right)^{1/n}  \right] \right]
\end{equation}

where the $\text{E}_m[x]$ function is defined by the integral

\begin{equation}
  \text{E}_m[x] =\int^{\infty}_1 \frac{e^{-xt} dt}{t^m}
\end{equation}

In this work, we have chosen to use $n=4$ consistently in all cases, which corresponds to employing the De Vaucouleurs profile. This profile is widely regarded as the best fit for describing the morphology of a central bulge. Taking this into account, the typical values in the literature \citep{WU20, CAR99} for $\Sigma_{e}$ can range from $10^{11}$ to $10^{14} M_\odot/\text{Mpc}^2$ and $R_e$ can vary from a few kpc to tens of kpc.

\subsubsection{The Burkert profile}
The Burkert profile \citep[][]{BUR95} is a phenomenological model used to describe the distribution of DM in astronomical objects. It specifically fits the cores in dwarf galaxy rotation curves, unlike the NFW profile, which predicts cuspy central density distributions. \citep{SAL00}. It is described by the equation

 \begin{equation}
     \rho_{\text{Burkert}}(r)=\frac{\rho_0 r_c^3}{\left(r+r_c\right)\left(r^2+r_c^2\right)},
 \end{equation}
where $\rho_0$ and $r_c$ are the central core density and core radius, respectively. The typical values in the literature \citep{GAM21,KAR14,KAR15} for $r_c$ are in the range of 1 to 100 kpc for galaxies and 10 to 1000 kpc for clusters. The central density of Burkert profiles, $\rho_0$, typically falls in the range of $10^{15} - 10^{18} M_{\odot}/\text{Mpc}^3$. In this case, the absence of an analytical expression for the convergence compels us to 
employ a numerical solution.

\section{ZOU clusters central positions}

\begin{figure*}[htp]
 \centering
 \includegraphics[width=0.49\textwidth]{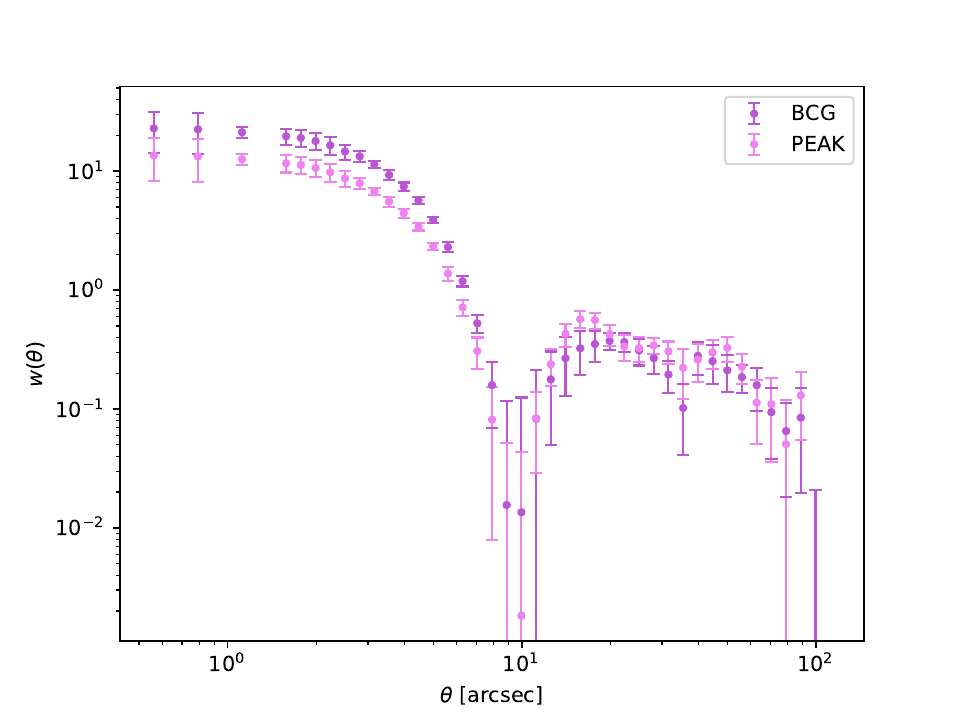}
 \includegraphics[width=0.49\textwidth]{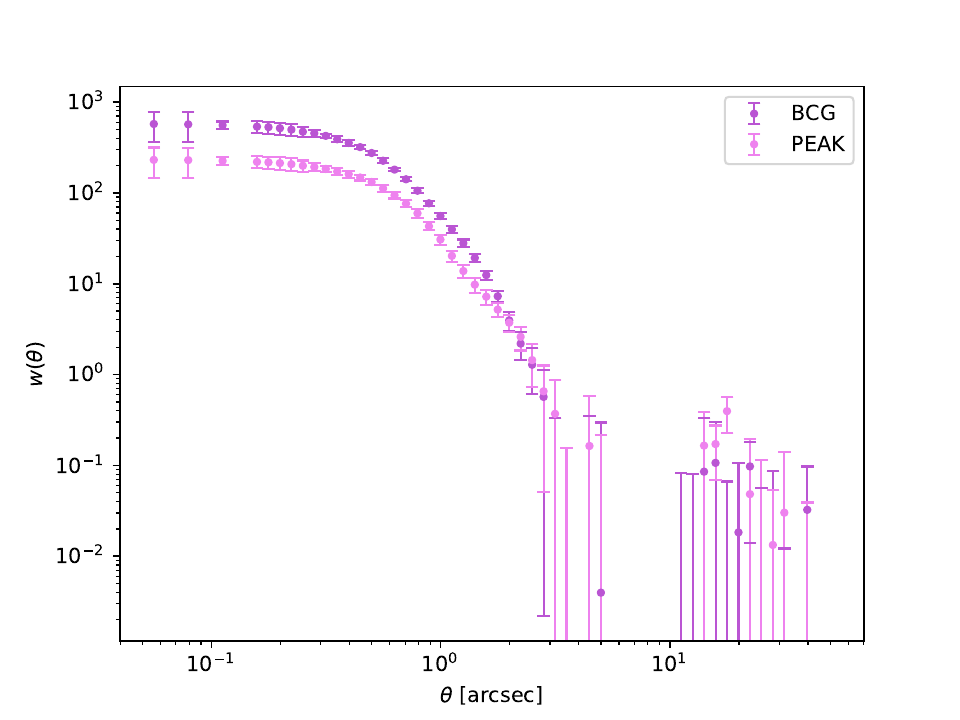}
 \caption{Cross-correlation data obtained with stacking using the background sample derived from the WISE catalogue and the foreground lens sample from the ZOU cluster catalogue. The purple data make use of the BCG coordinates to determine the clusters positions. The pink data were obtained with the cluster positions given by the peak coordinates of the luminosity distribution. Left panel: The case with npix=400, a pixel size of 0.5 arcsec, and $\sigma=2.4$ arcsec. Right panel: The case with npix=2000, a pixel size of 0.05 arcsec, and $\sigma=0.3$ arcsec.}
 \label{fig:cls_ZOU_PEAK_npix2000_s0.3}
\end{figure*}

When analysing the ZOU clusters case, we have two possible options for the central positions of the clusters. We ere provided with both the luminosity peak position (the peak coordinates of the luminosity distribution of the cluster) and the brightest galaxy positions (the coordinates of the brightest galaxy in the cluster). In order to estimate the cross-correlation signal we relied on position measurements, and therefore we needed to check if there were any significant variation in using one or anoter of the central positions of the clusters. We computed the cross-correlation using the stacking for the two different positional sets of coordinates. In particular, as is shown in Fig. \ref{fig:cls_ZOU_PEAK_npix2000_s0.3}, we compared the results with BCG and peak positions (purple and pink, respectively) for the npix=400 case (left case) and the npix=2000 case (right case). The background sample is the one from the WISE selection. Above 10 arcsec, both measurements agree, but the PEAK data are slightly lower than the BCG at angular scales below 10 arcsec.

\end{document}